\documentclass[useAMS,usenatbib,letterpaper]{mnras}

\usepackage{epsfig}
\usepackage{amsmath}
\usepackage{natbib}
\usepackage{aas_macros,amssymb}
\usepackage{tikz}

\begin{document}

\title[A Recipe for $P_{A^*}(k)$]{Predicting the Sufficient-Statistics Power Spectrum for Galaxy Surveys: A Recipe for $P_{A^*}(k)$}

\author[A. Repp \& I. Szapudi]{Andrew Repp\ \& Istv\'an Szapudi\\Institute for Astronomy, University of Hawaii, 2680 Woodlawn Drive, Honolulu, HI 96822, USA}


\label{firstpage}
\pagerange{\pageref{firstpage}--\pageref{lastpage}}
\maketitle

\begin{abstract}
Future galaxy surveys hope to realize significantly tighter constraints on various cosmological parameters. The higher number densities achieved by these surveys will allow them to probe the smaller scales affected by non-linear clustering. However, in these regimes, the standard power spectrum can extract only a portion of such surveys' cosmological information. In contrast, the alternate statistic $A^*$ has the potential to double these surveys' information return, provided one can predict the $A^*$-power spectrum for a given cosmology. Thus, in this work we provide a prescription for this power spectrum $P_{A^*}(k)$, finding that the prescription is typically accurate to about 5 per cent for near-concordance cosmologies. This prescription will thus allow us to multiply the information gained from surveys such as \textit{Euclid} and WFIRST.\\
\end{abstract}

\section{Introduction}
Cosmology -- the characterization of the Universe as a whole -- seeks a precise determination of the $\Lambda$CDM parameters. In particular, those dealing with dark energy and neutrino mass are not yet well-constrained.

Galaxy surveys constitute one of the most promising avenues of approach to this problem since they permit direct comparison between the observed galaxy distribution statistics and those predicted for various cosmological parameter values. The degree of anticipation associated with upcoming surveys such as \emph{Euclid} \citep{EuclidRedBook} and the Wide Field InfraRed Survey Telescope (WFIRST -- \citealp{WFIRST}) reflects the expected value of these surveys.

Of the various statistics one could use for this comparison, the power spectrum $P(k)$ of the overdensity $\delta = \rho/\overline{\rho} - 1$ (or, the two-point correlation function $\xi(r)$, which is its Fourier transform) has perhaps received the most attention (e.g. \citealp{Peebles1980,BaumgartFry1991,
Martinez2009}). One reason for this emphasis is that the two-point statistics of a Gaussian distribution completely characterize the distribution, and thus the power spectrum of the distribution exhausts the information inherent in it. And since the  fluctuations in the cosmic microwave background appear (so far) to be consistent with primordial Gaussianity \citep{Planck2015NG} -- and since the matter distribution remains roughly Gaussian on large, ``linear'' scales ($k \la 0.1h$ Mpc$^{-1}$) -- it follows that $P(k)$ is the statistic of choice for analyzing galaxy surveys at these scales.

Future surveys, however, promise a galaxy number density sufficient to probe much smaller scales. At these  scales, nonlinear gravitational amplification has over time produced an extremely non-Gaussian matter distribution (e.g., \citealp{FryPeebles1978,Sharp1984,SSB1992,Bouchet1993,
Gaztanaga1994}). The long positive tail -- and the correspondingly higher stochastic incidence of massive clusters -- heavily impacts the power spectrum on these small scales, resulting in large cosmic variance. This variance in turn markedly reduces the cosmological Fisher information captured by the power spectrum $P(k)$ (e.g., \citealp{RimesHamilton2005,RimesHamilton2006,NSR2006}). In particular, pushing a survey to smaller scales will not proportionately increase the Fisher information in $P(k)$, due to coupling between large and small Fourier modes \citep{MeiksinWhite1999,ScocZH1999}, which results in an ``information plateau'' \citep{NeyrinckSzapudi2007,LeePen2008,Carron2011, CarronNeyrinck2012,Wolk2013}. Hence, standard methods of analysis using the power spectrum can miss a large fraction -- in some cases, approximately half \citep{WCS2015a,WCS2015b,Repp2015} -- of the Fisher information inherent in these surveys.

The log transform provides a means of recovering this information \citep{Neyrinck2009}. Furthermore, using the theory of sufficient statistics (observables which capture all of the field's information) \citet{CarronSzapudi2013} find that for typical cosmological fields, the log transform yields an alternate statistic $A = \ln(1+\delta)$ which is essentially sufficient: i.e., this transformation counteracts nonlinear evolution to the point where the first two moments of $A$ contain virtually all of the cosmological information in any given survey pixel. It follows that the power spectrum $P_A(k)$ and mean $\langle A\rangle$ of this alternate statistic are the quantities one should study in order to deduce cosmological information from future surveys. To this end, \citet{ReppPAk} provide a simple and accurate fit for $P_A(k)$, and \citet{ReppApdf} provide a similar prescription for $\langle A\rangle$; they also show that a Generalized Extreme Value (GEV) model fits the  one-point distribution of $A$ quite well.

However, the statistic $A$ describes only the continuous dark matter distribution; the discreteness of galaxy counts (an empty cell of which would render the log transform problematic) requires modification of $A$. For such fields, \citet{CarronSzapudi2014} provide an analysis of the discrete optimal observable, denoting this observable as $A^*$. Hence, in order to avoid the information loss incurred by application of $P(k)$ to future dense galaxy surveys, one should perform the analysis using the $A^*$ statistic: i.e., one should compare the observed power spectrum $P_{A^*}(k)$ and mean $\langle A^* \rangle$ with the predictions of these quantities for various cosmological parameter values\footnote{We note that due to the nonlinear nature of the transformations, both $A$ and $A^*$ depend on pixel size in a way that $\delta$ does not. Defining the $A$ and $A^*$ statistics requires specification of the pixel scale, and evaluation of these statistics requires smoothing/binning to that scale.}.

To do so, of course, one requires the ability to make said predictions of $P_{A^*}(k)$ and $\langle A^* \rangle$. The aforementioned $A$-probability distribution allows prediction of $\langle A^*\rangle$ \citep{ReppAstarbias}, leaving characterization of the $A^*$-power spectrum the remaining problem. \citet{WCS2015b} identify the most salient feature of $P_{A^*}(k)$, namely, that it is biased with respect to the (continuous) log spectrum $P_A(k)$. \citet{ReppAstarbias} provide an a priori prescription for this bias in near-concordance cosmologies, with an accuracy better than 3 per cent for \emph{Euclid}-like surveys.

In this paper we complete the task begun in \citet{ReppAstarbias} by providing a detailed characterization of the $A^*$ power spectrum, including its discreteness plateau and the shape change incurred by passing from $A$ to $A^*$. We organize the work as follows: the relevant background appears in Section~\ref{sec:Astar}, which reviews and defines the $A^*$ statistic, and in Section~\ref{sec:bias}, which provides the $A^*$-bias prescription and briefly discusses its limits of applicability. Section~\ref{sec:disc} analyzes the plateau introduced into $P_{A^*}(k)$ by the discreteness of the galaxy field -- analogous to (but not equal to) the $1/\overline{n}$ shot noise plateau in the standard power spectrum. Section~\ref{sec:shape} then characterizes (and provides a prescription for) the shape of $P_{A^*}(k)$. We quantify the accuracy of our prescription in Section~\ref{sec:accuracy}, and we conclude in Section~\ref{sec:concl}.

\section{The Discrete Sufficient Statistic $A^*$}
\label{sec:Astar}
\citet{CarronSzapudi2013} demonstrate that the log transform $A = \ln (1+\delta)$ yields a statistic that is essentially ``sufficient,'' in that it extracts (virtually) all of the Fisher information in a survey.\footnote{Specifically, by this we mean that all of the information in the one-point distribution is contained in the first two moments of the pixel values after application of the transformation.} Because this transformation thus approximately Gaussianizes the overdensity field $\delta$, the power spectrum $P_A(k)$ of the log overdensity extracts substantially more information at small scales than the power spectrum $P(k)$ of the overdensity field itself.

In reality, of course, one surveys not the dark matter density but the galaxy distribution, thus introducing shot noise. Under the assumption that light traces mass, galaxy surveys represent a discretization of the underlying dark matter field. Since $A$ is a continuous variable, it requires modification in order to serve as an efficient information-extractor from a discrete field.

For this reason, \citet{CarronSzapudi2014} provide the appropriate generalization of the log transform to discrete fields, formulating a statistic which they denote $A^*$, and showing that it is a good approximation to a sufficient statistic for discrete fields. $A^*$ is the Bayesian reconstruction of the underlying dark matter field, given the measured galaxy counts $N$. In particular, to construct $A^*(N)$ one must first know the probability distribution $\mathcal{P}(A)$ of the log density contrast $A$ (or equivalently the distribution $\mathcal{P}(\delta)$ of $\delta$).\footnote{Throughout this article we distinguish probability distributions from power spectra by using script and roman letters, respectively: thus $\mathcal{P}(A)$, but $P_A(k)$.} One must also assume a discrete sampling scheme $\mathcal{P}(N|A)$, which provides the probability of finding $N$ galaxies given an underlying dark matter log density $A$. Perhaps the simplest such scheme is Poisson sampling, for which
\begin{equation}
\mathcal{P}(N|A) = \frac{1}{N!} \left( \overline{N} e^A\right)^N \exp \left( -\overline{N}e^A \right),
\label{eq:Poisson}
\end{equation}
where $\overline{N}$ is the average number of galaxies per survey pixel.

Given these two distributions $\mathcal{P}(A)$ and $\mathcal{P}(N|A)$, \citet{CarronSzapudi2014} define $A^*(N)$ as the value of $A$ which maximizes $\mathcal{P}(A)\mathcal{P}(N |A)$; they further show that $A^*(N)$ is also the peak of the Bayesian a posteriori distribution for the dark matter log density in a survey pixel containing $N$ galaxies.

In the following two subsections, we provide expressions for $A^*$ under the assumption of Poisson sampling, given two approximations for the distribution of dark matter $\mathcal{P}(A)$. It is straightforward to define $A^*$ for other dark matter probability distributions and sampling schemes.

\subsection{$A^*$ for a Lognormal Matter Distribution}
A lognormal model for the cosmic matter distribution arises naturally from simple assumptions \citep{ColesJones,KTS2001} and is an accurate approximation in the projected, two-dimensional case. It is this model (with Poisson sampling) with which \citet{CarronSzapudi2014} explicitly deal, concluding that $A^*(N)$ is the solution of
\begin{equation}
e^{A^*} + \frac{A^*(N)}{\overline{N} \sigma_A^2} = \frac{N-1/2}{\overline{N}}.
\label{eq:Astarln}
\end{equation}
Here $\overline{N}$ is the average number of galaxies per survey pixel and $\sigma^2_A$ is the variance of the log dark matter density contrast. It is through these two parameters, respectively, that $A^*$ depends on the discrete sampling scheme and the underlying dark matter distribution, respectively. 

\subsection{$A^*$ for a log-GEV Matter Distribution}

If we consider the matter distribution in three dimensions (rather than projecting to two), it departs significantly from the lognormal on translinear scales. We show in \citet{ReppApdf} that a Generalized Extreme Value (GEV) distribution provides a better fit to the $A$-distribution (in contrast to a Gaussian distribution for $A$, which would follow from a lognormal distribution for $\delta$). In particular, we show that for redshifts $z=0$ to 2 and for scales down to $2h^{-1}$Mpc, the following distribution is an excellent fit to the Millennium Simulation \citep{MillSim} results:
\begin{equation}
\label{eq:GEV}
\mathcal{P}(A) = \frac{1}{\sigma_G} t(A)^{1+\xi_G} e^{-t(A)},
\end{equation}
where
\begin{equation}
\label{eq:GEV_t}
t(A) = \left(1 + \frac{A - \mu_G}{\sigma_G}\xi_G\right)^{-1/\xi_G}.
\end{equation}
Here, $\mu_G$, $\sigma_G$, and $\xi_G$ depend on the mean $\langle A \rangle$, variance $\sigma_A^2$, and skewness $\gamma_1$ of $A$, as follows:
\begin{equation}
\gamma_1 = -\frac{\Gamma(1-3\xi_G) - 3\Gamma(1-\xi_G)\Gamma(1-2\xi_G) + 2\Gamma^3(1-\xi_G)}{\left(\Gamma(1-2\xi_G) - \Gamma^2(1-\xi_G)\right)^{3/2}}
\label{eq:xiG}
\end{equation}
\begin{equation}
\sigma_G = \sigma_A \xi_G \cdot \left(\Gamma(1-2\xi_G) - \Gamma^2(1-\xi_G)\right)^{-1/2}
\label{eq:sigG}
\end{equation}
\begin{equation}
\mu_G = \langle A \rangle - \sigma_G \frac{\Gamma(1-\xi_G) - 1}{\xi_G},
\label{eq:muG}
\end{equation}
where $\Gamma(x)$ is the gamma function.

In \citet{ReppAstarbias} we show that Poisson sampling of a GEV distribution yields the following equation for $A^*$:
\begin{multline}
\label{eq:AstarGEV}
\frac{1}{\sigma_G} \left( 1 + \frac{A^*(N) - \mu_G}{\sigma_G} \xi_G \right)^{-1-\frac{1}{\xi_G}} + N \\
= \frac{1+\xi_G}{\sigma_G + \left(A^*(N) - \mu_G\right)\xi_G} + \overline{N}e^{A^*(N)}.
\end{multline}
Once again, $A^*(N)$ depends on the sampling scheme through the $\overline{N}$ parameter, and it depends on the dark matter distribution through the $\mu_G$, $\sigma_G$, and $\xi_G$ parameters.

It is Equation~\ref{eq:AstarGEV} which we use for calculating $A^*$ throughout the remainder of this paper.

\section{The Bias of the $A^*$-Power Spectrum}
\label{sec:bias}

To a first approximation, the power spectrum of $A^*$ exhibits the same shape as its continuous analog $P_A(k)$, with the exception of a multiplicative bias \citep{WCS2015b}. \citeauthor{WCS2015b} also provide an approximate formula for this bias in the case of a two-dimensional (projected) galaxy survey, assuming a lognormal probability distribution.

For conceptual clarity, it is important to note that this bias (denoted $b^2_{A^*}$ below) is unrelated to the more commonly encountered ``galaxy bias,'' which expresses the fact that galaxies  cluster more strongly than dark matter. The latter depends on galaxy formation physics, whereas the former is a statistical effect of the passage from $A$ to $A^*$. Even in the case of identical galaxy- and dark matter-clustering (i.e., a galaxy bias of unity), the power spectrum $P_{A^*}(k)$ would nevertheless exhibit an offset from $P_A(k)$ in the amount of $b^2_{A^*}$. In this work we deal solely with the $A^*$-bias; we do not again mention galaxy bias until the penultimate sentence of Section~\ref{sec:concl}.

To deal with the full three-dimensional data, \citet{ReppAstarbias} derive an expression for the bias in terms of the discrete sampling scheme $\mathcal{P}(N|A)$ and the underlying dark matter distribution $\mathcal{P}(A)$:
\begin{equation}
b^2_{A^*} = \frac{1}{\sigma_A^4} \left\lbrace \sum_N\int dA\,(A-\overline{A})(A^* - \overline{A^*})\mathcal{P}(N|A)\mathcal{P}(A) \right\rbrace^2.
\label{eq:Astarbias}
\end{equation}
The accuracy of this formula depends on the assumption that at large scales the correlation functions $\xi$ of $A$ and of $A^*$ have the same shape, so that $\xi_{A^*}(r) = b_{A^*}^2 \xi_A(r)$. This assumption is not completely valid, as we mention below (albeit in the context of the power spectra rather than the correlation functions); indeed, when the average number $\overline{N}$ of particles per cell is too low ($\overline{N} \la 0.5$), the shapes are sufficiently different that Equation~\ref{eq:Astarbias} yields too low a value for $b^2_{A^*}$. However, the practical applicability of galaxy survey results is limited to scales at which $\overline{N} \ga 1$, and in this regime we can use Equation~\ref{eq:Astarbias} to provide the overall bias and then make the slight shape modifications discussed in the following sections.

Note that below (see Section~\ref{sec:shape}) we refine our understanding of this bias in terms of the decomposition of $A^*$ accomplished in Section~\ref{sec:disc}. Equation~\ref{eq:Astarbias} receives similar modification.

\section{Discreteness Effects in $P_{A^*}(k)$}
\label{sec:disc}

It is well-known (e.g., \citealp{Peebles1980}) that if one Poisson-samples a continuous density contrast field $\delta(\mathbf{r})$ to obtain a discrete density contrast $\delta_d(\mathbf{r})$, then the power spectra of the two fields relate as follows: 
\begin{equation}
P_d(k) = P(k) + \frac{1}{\overline{n}},
\label{eq:1_over_n}
\end{equation}
where $P_d(k)$ is the power spectrum of $\delta_d$, $P(k)$ is the power spectrum of $\delta$, and $\overline{n}$ is the number density in units of inverse volume. As \citet{Neyrinck2011} note, the discrete log spectrum exhibits a similar plateau at high values of $k$.

Thus, we here derive an expression for this analogous $A^*$-discreteness plateau. We first (Section~\ref{sec:N}) derive the power spectrum for the field of galaxy number counts $N$ rather than for the density contrast $\delta_d$. Then we derive (Section~\ref{sec:f}) the power spectrum for an arbitrary function $f$ of $N$, under the assumption that the field $N$ is uncorrelated. We next (Section~\ref{sec:deltaAstar}) decompose $A^*$ into correlated and uncorrelated components, permitting determination of the $A^*$-discreteness plateau. Discussion of the plateau follows in Section~\ref{sec:discretediscussion}.

\subsection{The Number Count Field}
\label{sec:N}
To derive the discreteness plateau for $A^*$, we first consider the power spectrum $P_N(k)$ for the actual number count field (i.e., the number of galaxies in each cell). Number counts depend on survey cell size in a way that densities do not. In any given survey cell $\mathbf{r}_i$, we have the number count $N_i = \overline{N} ( \delta_d(\mathbf{r}_i) + 1)$, where $\overline{N}$ is the mean number of counts per cell. Thus to obtain $P(N)$ we simply multiply Equation~\ref{eq:1_over_n} by the square of $\overline{N}$:
\begin{equation}
P_N(k) = \overline{N}^{\,2} P(k) + \frac{\overline{N}^{\,2}}{\overline{n}} = \overline{N}^{\,2} P(k) + \overline{N}\, \delta V,
\label{eq:PN}
\end{equation}
where $\delta V$ is the size of a survey cell.

\subsection{Transforming Uncorrelated Counts}
\label{sec:f}
In practice we smooth (integrate) over finite subcells -- the survey pixels -- before taking the power spectrum. Since integration is a linear operation, the only effect of doing so is the introduction of pixel window effects into the continuous (correlated) part of Equations~\ref{eq:1_over_n} and \ref{eq:PN}; the discreteness term is unaffected.

The situation changes with the $A^*$ field. If we begin with a discrete field $n(\mathbf{r})$, we would integrate over a finite subcell to obtain $N_i$ and then determine $A^*(N_i)$. Only after this highly-nonlinear $A^*$ transformation do we we take the Fourier transform to get $P_{A^*}(k)$.

To address this problem of nonlinearity, we begin by considering an uncorrelated field of number counts $N$, and we let $f(N)$ be an arbitrary transformation of this field, subject only to the condition that the transformed field $f(N)$ remain uncorrelated. 

We first note that the power spectrum of any uncorrelated field is constant. This being the case, let $C$ be the constant value of the power spectrum $P_f(k)$ of $f(N)$, and consider a cubical volume divided into cubical survey cells of side length $(\delta V)^{1/3}$. Then $k_N = (\delta V)^{-1/3}\pi$ is the Nyquist frequency of the survey, and we can obtain the variance $\sigma_f^2$ of the field by integrating the power spectrum over a cube in $k$-space of side length $2k_N$: 
\begin{equation*}
\sigma_f^2 = \int_{-k_N}^{k_N} \frac{dk_1}{2\pi} \int_{-k_N}^{k_N} \frac{dk_2}{2\pi} \int_{-k_N}^{k_N} \frac{dk_3}{2\pi} P_f(k) = C\left( \frac{k_N}{\pi} \right)^3.
\end{equation*}
But $k_N^3 = \pi^3/\delta V$, so $\sigma_f^2 = C/\delta V$. We thus obtain the simple result that
\begin{equation}
P_f(k) = \delta V \cdot \sigma_f^2.
\label{eq:PAsdisc}
\end{equation}

We can gain more insight into this result by temporarily imposing the additional assumption that the transformation $f$ is a function $f(N)$ of only the number counts in each cell (rather than, say, depending on the underlying dark matter density).

In this case, if we assume that $\overline{N}$ is small enough that $N_i = 1$ or 0 for all cells, then the transformation is linear, being determined by the two points $N = 0 \longmapsto f(0)$ and $N = 1 \longmapsto f(1)$. The transformation from $N$ to $f$ then consists of only a scaling (and an irrelevant amplitude shift). Thus the discreteness correction in Equation~\ref{eq:PN} transforms to
\begin{equation}
P_f(k) = (f(1) - f(0))^2 \,\overline{N} \delta V, \mbox{\hspace{0.5cm}}\overline{N} \ll 1.
\label{eq:lowN}
\end{equation}

We can then deal in turn with arbitrary number densities by recalling that the $f$-variances of two transformed fields of number counts will be the integrals of their (constant) power spectra, and thus the variances will be in the same ratio as the values of the $f$-power spectra:
\begin{equation}
  \frac{\sigma_{f,\overline{N}_1}^2}{\sigma_{f,\overline{N}_2}^2} = \frac{P_{f,\overline{N}_1}(k)}{P_{f,\overline{N}_2}(k)}.
\label{eq:varratios}
\end{equation}

Let us denote the quantities in the low-$\overline{N}$ limit as $\overline{N}_0$, $\sigma_{f_0}^2$, etc.; in this limit Equation~\ref{eq:lowN} holds. Then for any $\overline{N}$, Equation~\ref{eq:varratios} allows us to write
\begin{equation}
P_f(k) = \sigma_f^2 \frac{P_{f_0}(k)}{\sigma^2_{f_0}} = \sigma_f^2 \frac{\overline{N}_0 \left(f(1) - f(0)\right)^2 \delta V}{\sigma_{f_0}^2}.
\label{eq:gen_shotnoise}
\end{equation}
Working to first order in $\overline{N}_0$, we can say that the probability of one particle in the cell is $\mathcal{P}(1) = \overline{N}_0$ and the probability of an empty cell is $\mathcal{P}(0) = 1 - \overline{N}_0$. In this limit,
\begin{align*}
\sigma_{f_0}^2 = {} & \langle f^2 \rangle_0 - \langle f \rangle^2_0 \\
   = {} &(\mathcal{P}(0)\cdot f(0)^2 + \mathcal{P}(1) \cdot f(1)^2 ) \\
   & - (\mathcal{P}(0)\cdot f(0) + \mathcal{P}(1) \cdot f(1) )^2\\
   = {} & ((1 - \overline{N}_0) f(0)^2 + \overline{N}_0 \cdot f(1)^2 ) \\
   & - ((1 - \overline{N}_0)\cdot f(0) + \overline{N}_0 \cdot f(1) )^2\\
   = {} & \overline{N}_0 \left(f(1) - f(0)\right)^2,
\end{align*}
and Equation~\ref{eq:PAsdisc} follows.

However, our original derivation of Equation~\ref{eq:PAsdisc} does not require $f$ to be a function of only number counts $N$; rather, it simply requires that the transformed field $f$ be uncorrelated. Hence, the transformation $f$ can depend on the dark matter density, as long as that dependence does not introduce correlations into the transformed field.

\subsection{The Discreteness Plateau for $A^*$}
\label{sec:deltaAstar}
To extend Equation~\ref{eq:PAsdisc} to $A^*$ on correlated fields, we note that there are two sources of variation in $A^*$: the value of  $A^*\!(\mathbf{r_1})$ might differ from that of  $A^*\!(\mathbf{r_2})$ because the underlying dark matter densities differ (i.e., $A(\mathbf{r_1}) \neq A(\mathbf{r_2})$); or it might differ because of stochasticity during discrete sampling.

Equivalently, there are two effects involved in the passage from $A$ to $A^*$. First, the mapping in Equation~\ref{eq:AstarGEV} (as well as that in Equation~\ref{eq:Astarln}) is inherently nonlinear in $A$. Second, in addition to this nonlinearity we have the stochastic nature of the discrete sampling process, reflected in the fact that $A^*$ is a function of $N$ rather than of $A$.

\begin{figure}
\leavevmode\epsfxsize=9cm\epsfbox{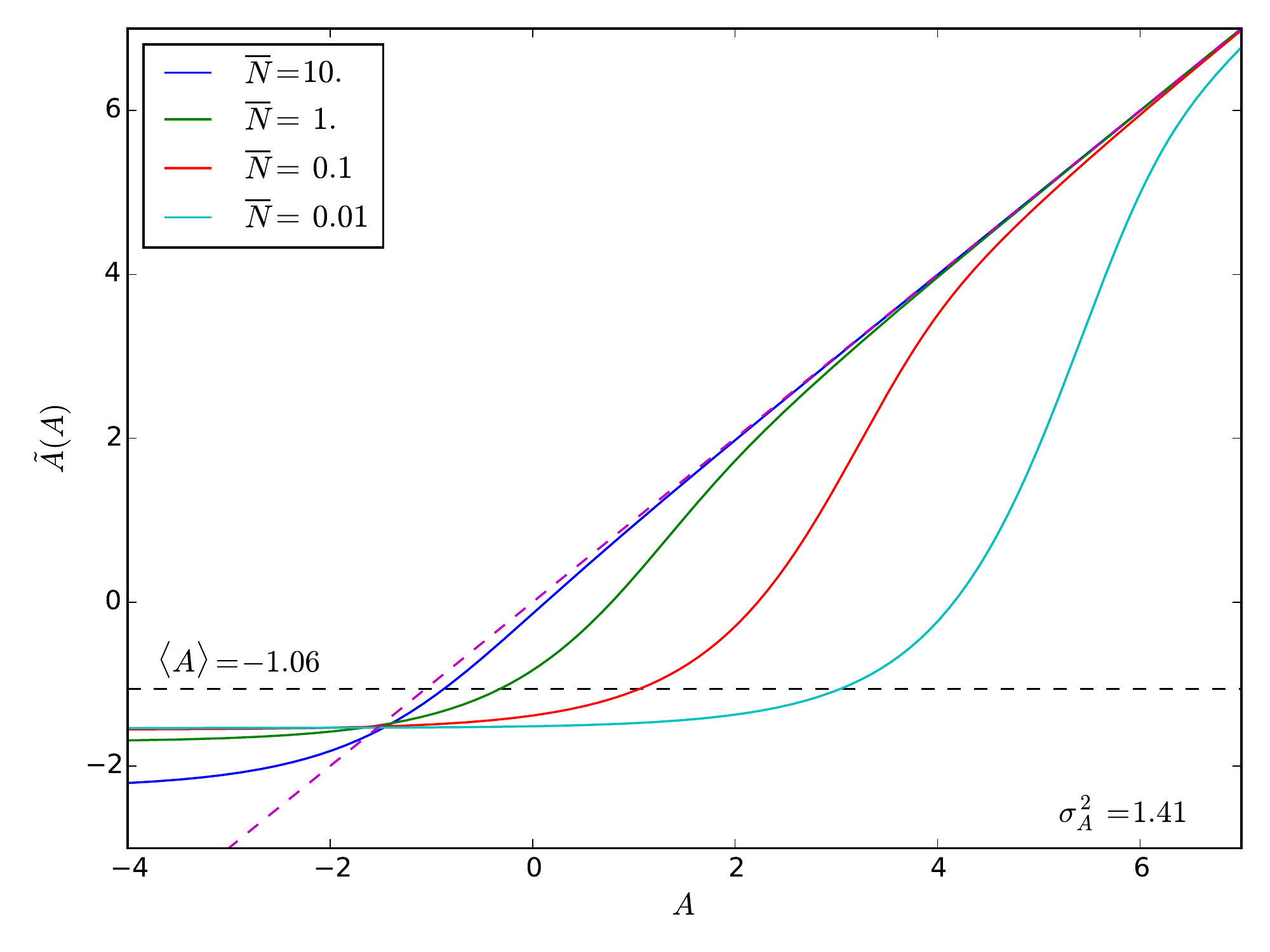}
\caption{The continuous component $\tilde{A}(A)$ of $A^*$, for several values of mean galaxies per cell $\overline{N}$. The dashed magenta line marks $\tilde{A} = A$, and the dashed black line marks the mean value of $A$. The calculation assumes a Planck 2015 \citep{Planck2015} cosmology at $z = 0$, with cubical survey cells of side length $1.95h$ Mpc$^{-1}$.}
\label{fig:Atildeplot}
\end{figure}

In order to disentangle these effects, we decompose $A^*\!(\mathbf{r})$ into two components. The first component is the expected value of $A^*$ given an underlying (dark matter) value of $A(\mathbf{r})$; we denote this component $\tilde{A}$:
\begin{equation}
\tilde{A}\left(A(\mathbf{r})\right) \equiv \langle A^* \rangle \Big|_A = \sum_{N=0}^\infty \mathcal{P}(N|A) A^*(N).
\label{eq:Atilde}
\end{equation}
$\tilde{A}(A)$ thus encapsulates the nonlinearity of $A^*$ without its stochasticity.

Figure~\ref{fig:Atildeplot} shows $\tilde{A}(A)$ for various values of $\overline{N}$. Inspection of this figure shows that $\tilde{A}$ approaches $A$ for large values of $A$, as expected, since for high dark matter densities the effects of discretization are increasingly irrelevant. Likewise, the higher the value of $\overline{N}$, the less difference there is between $\tilde{A}(A)$ and $A$ itself.

On the other hand, we see that $\tilde{A}$ asymptotes to a distinct minimum value (which depends on $\overline{N}$); this minimum corresponds to the value of $A$ at which $\mathcal{P}(N\!=\!0\,|A) \approx 1$, so that $\tilde{A} \approx A^*\!(N\!=\!0)$. Again as expected, this minimum value of $\tilde{A}$ decreases with $\overline{N}$, because with more galaxies, one obtains better resolution in low-density regions. The fact that the minimum $\tilde{A}$ consistently falls below $\langle A \rangle$ -- even for very small $\overline{N}$ -- is a result of the fact that the most likely value of $A$ is less than the mean of $A$ (because of the positive skewness of the $A$-distribution).

Thus far in this section we have considered the ``continuous'' component of $A^*$; we now turn to the remaining component, which we denote $\delta\!A^*$: 
\begin{equation}
\delta\!A^*(\mathbf{r}) \equiv A^*(\mathbf{r}) - \tilde{A}\left(A(\mathbf{r})\right).
\end{equation}
This component contains the stochasticity induced by discreteness at a particular point in the field.

Since $\delta\!A^*$ is the result solely of stochasticity in the Poisson sampling, it is reasonable to assume that the field $\delta\!A^*\!(\mathbf{r})$ is uncorrelated -- a fact which we demonstrate rigorously in the Appendix. And since $\delta\!A^*$ depends on number counts, we can use Equation~\ref{eq:PAsdisc} with the transformation $f:N(\mathbf{r}) \longrightarrow \delta\!A^* (\mathbf{r})$ and obtain the power spectrum of $\delta\!A^*$:
\begin{equation}
P_{\delta\!A^*} (k) = \delta V \cdot \sigma^2_{\delta\!A^*}.
\label{eq:PkdeltAstar}
\end{equation}

This (constant) value gives the discreteness plateau of the $A^*$-power spectrum, and we can write
\begin{equation}
P_{A^*}(k) = P_{\tilde{A}}(k) + \delta V \cdot \sigma^2_{\delta\!A^*}.
\end{equation}
In addition, the Appendix also shows that $\sigma^2_{\delta\!A^*} = \sigma^2_{A^*} - \sigma^2_{\tilde{A}}$; thus we conclude
\begin{equation}
P_{A^*}(k) = P_{\tilde{A}}(k) + \delta V \cdot \left(\sigma^2_{A^*} - \sigma^2_{\tilde{A}}\right).
\label{eq:PAswithPdeltAs}
\end{equation}

Using the probability distributions to explicitly write the variances from Equation~\ref{eq:PAswithPdeltAs}, we have
\begin{align}
\sigma^2_{A^*} & = \sum_N \int dA\,\,\mathcal{P}(A) \mathcal{P}(N|A)\left(A^*(N) - \langle A^* \rangle \right)^2 \label{eq:sig2Astar}\\
\sigma^2_{\tilde{A}} & = \int dA\,\,\mathcal{P}(A) \left(\tilde{A}(A) - \langle \tilde{A} \rangle \right)^2 \label{eq:sig2Atilde}.
\end{align}
Likewise, one can express the means $\langle A^* \rangle$ and $\langle \tilde{A} \rangle$ as moments of the appropriate probability distributions.

Note that it is straightforward to verify that we recover the standard $1/\overline{n}$ discreteness plateau by formally setting $A^*(N) = \delta_d = N/\overline{N} - 1$, $\tilde{A}(A) = \delta = e^A - 1$, and $\mathcal{P}(N|A) = \mathrm{Pois}(\overline{N}e^A)$ in Equations~\ref{eq:PAswithPdeltAs}, \ref{eq:sig2Astar}, and \ref{eq:sig2Atilde}, where $\mathrm{Pois}(\lambda)$ denotes a Poisson distribution with mean $\lambda$.

\begin{figure*}
\leavevmode\epsfxsize=18cm\epsfbox{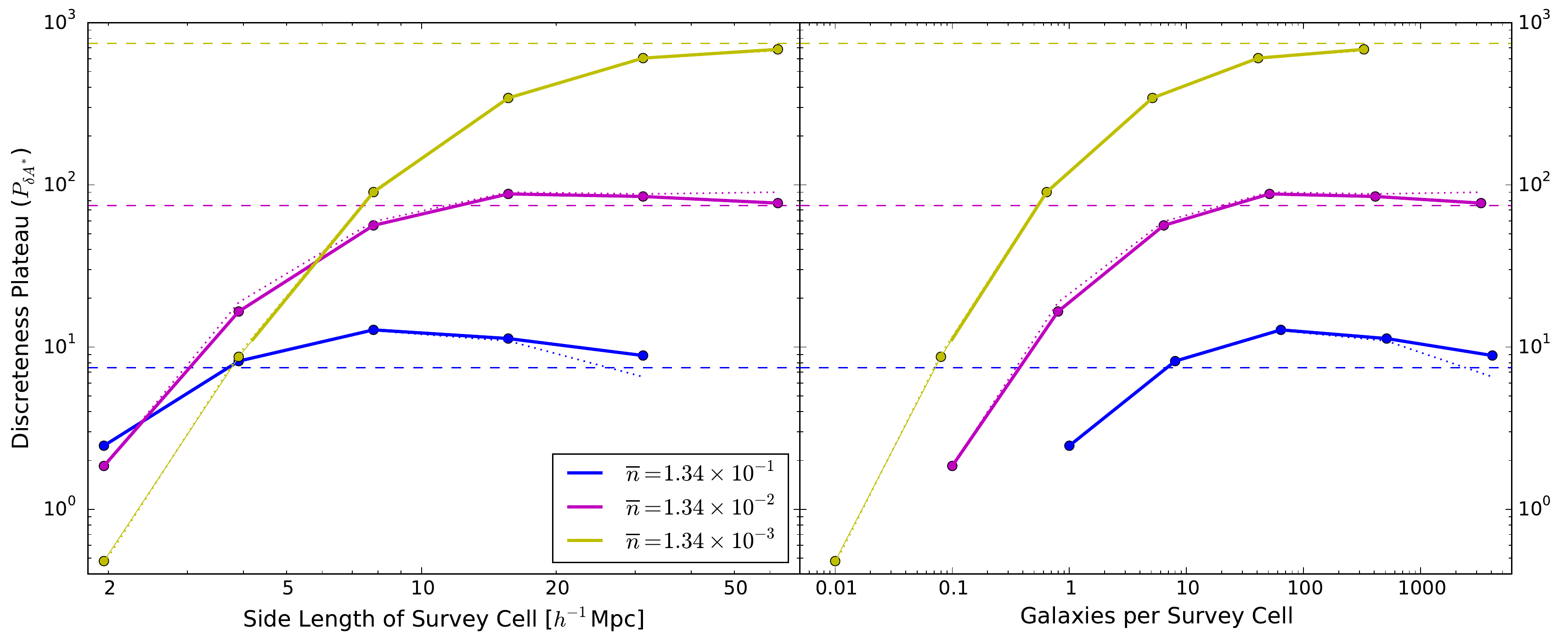}
\caption{Typical values of the discreteness plateau of the $A^*$-power spectrum, for three different number densities (in units of $h^3$Mpc$^{-3}$) at scales ranging from $2$--$62h^{-1}$ Mpc, in terms of pixel side length (left panel) and of galaxies per pixel (right panel). The number densities correspond to 0.01, 0.1, and 1 galaxy per cell at the smallest scale shown. Solid curves show the results using our GEV fit for the log density distribution \citep{ReppApdf}; dotted curves show the results using discrete realizations of the Millennium Simulation dark matter field. The dashed lines show the $1/\overline{n}$ discreteness plateau for Poisson noise. For the lowest number density
($1.34\times 10^{-3}h^3$Mpc$^{-3}$), the transition from a thick to a thin line marks the Poisson-noise limit at which $P(k) < 1/\overline{n}$. The calculations assume the cosmology of the original Millennium Simulation \citep{MillSim} at $z = 0$, with cubical survey cells of side length $1.95h$ Mpc$^{-1}$.}
\label{fig:PkdeltAstar}
\end{figure*}

\subsection{Discussion}
\label{sec:discretediscussion}
Figure~\ref{fig:PkdeltAstar} displays typical values for this discreteness plateau $P_{\delta\!A^*}$ for three galaxy number densities and for a variety of pixel side lengths. We display both the results of using our GEV probability distribution and those of applying Equation~\ref{eq:PkdeltAstar} to discrete realizations of the Millennium Simulation. In Sections~\ref{sec:shape} and \ref{sec:accuracy} we further describe these discrete realizations, and we also discuss the (slight) disparity between the two sets of calculations. At this point, however, we note a few general trends.

First, at large scales the discreteness plateau of $P_{A^*}(k)$ approaches the standard $1/\overline{n}$ value for the galaxy spectrum $P_g(k)$. This behavior is not unexpected, since at large scales the density contrast is small, so that $A = \ln(1 + \delta) \approx \delta$; since $A^*$ is the discrete analog of $\delta_g$, it is unsurprising that their behaviors match on these scales. At these scales, the level of the plateau increases as number density decreases.

Second, it is interesting that the approach to this $1/\overline{n}$ value is not necessarily monotonic: there are scales and number densities at which $P_{\delta\!A^*}$ is slightly higher than $1/\overline{n}$, whereas $P_{A^*}(k)$ is in general lower than $P(k)$ (because of the bias -- see Section~\ref{sec:bias}). However, because $P_{\delta\!A^*}$ exhibits some cosmology-dependence (through the effects of the probability distribution $\mathcal{P}(A)$ on $\sigma_{A^*}^2$ and $\sigma^2_{\tilde{A}}$), it is still possible to extract information on scales at which $P_{\delta\!A^*}$ dominates over $P_{\tilde{A}}(k)$.

Finally, we see (left panel of Figure~\ref{fig:PkdeltAstar}) that at the smallest scales the relationship between number density and $P_{\delta\!A^*}$ is inverted (with respect to the relationship at large scales) -- namely, lower number densities imply a lower plateau. At first this reversal appears counterintuitive -- why would lower number densities effectively produce \emph{less} shot noise?

But this behavior is a direct consequence of the scale-dependent nature of the map $A^*\!(N)$, which depends explicitly on counts per cell $\overline{N}$ (rather than $\overline{n}$, by which we denote counts per unit volume). The $A^*$ map itself thus depends on the smoothing scale, in contrast to the galaxy overdensity $\delta_g(N) = N/\overline{N} -1$, in which the cell volume affects $N$ and $\overline{N}$ equally. As a result, the galaxy power spectrum $P_g(k)$ does not depend on the smoothing scale (except for pixelation effects, etc.), whereas the power spectrum of the nonlinear map $A^*$ does.

For this reason, the standard derivation of the shot noise plateau for $P_g(k)$ proceeds by subdivision of cells until each cell contains either $N=0$ or $N=1$, and this procedure is permissible because $P_g(k)$ is independent of the pixel size. The resulting plateau at $1/\overline{n}$ essentially yields the average volume containing a single galaxy, and it is this average volume which determines the level of the plateau.

However, when we attempt to apply the same procedure to $P_{A^*}(k)$, we find that the subdivision process changes the $A^*$ map itself; it thus also changes the spectrum and hence the spectrum's shot-noise plateau. It is for this reason that the curves in Figure~\ref{fig:PkdeltAstar} are not straight lines (and also for this reason that Equation~\ref{eq:lowN} required modification to Equation~\ref{eq:gen_shotnoise}).

It follows that the left-hand panel of Figure~\ref{fig:PkdeltAstar} displays two entangled effects: first, there is the average volume per galaxy (i.e., the number density $\overline{n}$), which sets the plateau for the standard spectrum and is independent of pixel size. Second, there is the effect of the number of galaxies per cell ($\overline{N}$) on the $A^*$-map -- and this $\overline{N}$ depends on both cell size and $\overline{n}$. For instance, in the left-hand panel of Figure~\ref{fig:PkdeltAstar}, cells with sides of length $5h^{-1}$ Mpc correspond to $\overline{N}$-values ranging from 0.17 to 17, depending on $\overline{n}$.

To disentangle these two effects, we can plot the discreteness plateau levels as functions of $\overline{N}$ rather than scale (as in the right-hand panel of Figure~\ref{fig:PkdeltAstar}). Since any given $\overline{N}$ yields the same $A^*$-map\footnote{This statement is only approximately true, since the distribution $\mathcal{P}(A)$ also depends on the smoothing scale.}, the difference in the three curves is due only to the difference in number densities (or equivalently, due only to the average volume per galaxy). In terms of Equation~\ref{eq:PkdeltAstar}, the right-hand panel keeps $\overline{N}$ constant and thus forces $\sigma^2_{\delta \!A^*} = (\sigma^2_{A^*} - \sigma^2_{\tilde{A}})$ to be (relatively) constant, given the (relative) constancy of the $A^*$ map. However, maintaining a constant $\overline{N}$ requires varying pixel volumes $\delta V$ for varying number densities $\overline{n}$, and it is this $\delta V$ which is analogous to the $1/\overline{n}$ value in the standard shot noise plateau. When we compare the $A^*$-plateau levels in this way we see that, as expected, it is the higher number densities which correspond to lower plateaus. On the other hand, if we insist on comparison at a constant spatial scale (as in the left-hand panel), then we force a constant $\delta V$, and the variation in the $A^*$-map for different values of $\overline{N}$ causes the intersecting curves in that panel.

\section{The Shape of the $A^*$ Spectrum}
\label{sec:shape}

\subsection{Parametrizing $\tilde{A}$}
We now turn to more accurate characterization of the shape of $P_{A^*}(k)$. In Section~\ref{sec:disc} we decomposed $A^*$ into the continous nonlinear map $\tilde{A}(A)$ and the stochastic component $\delta\!A^*$, which in turn allowed us to decompose the power spectrum as in Equation~\ref{eq:PAswithPdeltAs}.

It follows that the $A^*$-bias from Section~\ref{sec:bias} belongs, strictly speaking, to the continuous component $\tilde{A}$, since the stochastic part of $A^*$ introduces only an additive constant to its power spectrum. Thus, the same procedure  used in \citet{ReppAstarbias} allows us to write the bias formula from Equation~\ref{eq:Astarbias} in terms of $\tilde{A}$ rather than $A^*$:
\begin{equation}
b^2_{A^*} = \frac{1}{\sigma_A^4} \left\lbrace \int dA\,(A-\overline{A})\left(\tilde{A}(A) - \langle\tilde{A}(A) \rangle\right)\mathcal{P}(A) \right\rbrace^2,
\label{eq:Atildebias}
\end{equation}
so that we now write
\begin{equation}
P_{A^*}(k) \approx b^2_{A^*} P_A(k) + \delta V \cdot \left(\sigma^2_{A^*} - \sigma^2_{\tilde{A}}\right).
\label{eq:prelimPAs}
\end{equation}

This representation provides the correct overall bias and discreteness plateau. However, the nonlinear nature of the $\tilde{A}$ transformation also introduces slight but non-negligible changes into the shape of $P_{\tilde{A}}(k)$. As long as the number of particles per cell is not too low ($\overline{N} \ga 0.5$), we find that it will suffice to correct the bias in Equation~\ref{eq:prelimPAs} with two shape-change terms. (As we mention in Section~\ref{sec:bias}, at lower number densities the shape change is severe enough to render Equation~\ref{eq:Atildebias} inaccurate.)

\begin{figure*}
\leavevmode\epsfxsize=18cm\epsfbox{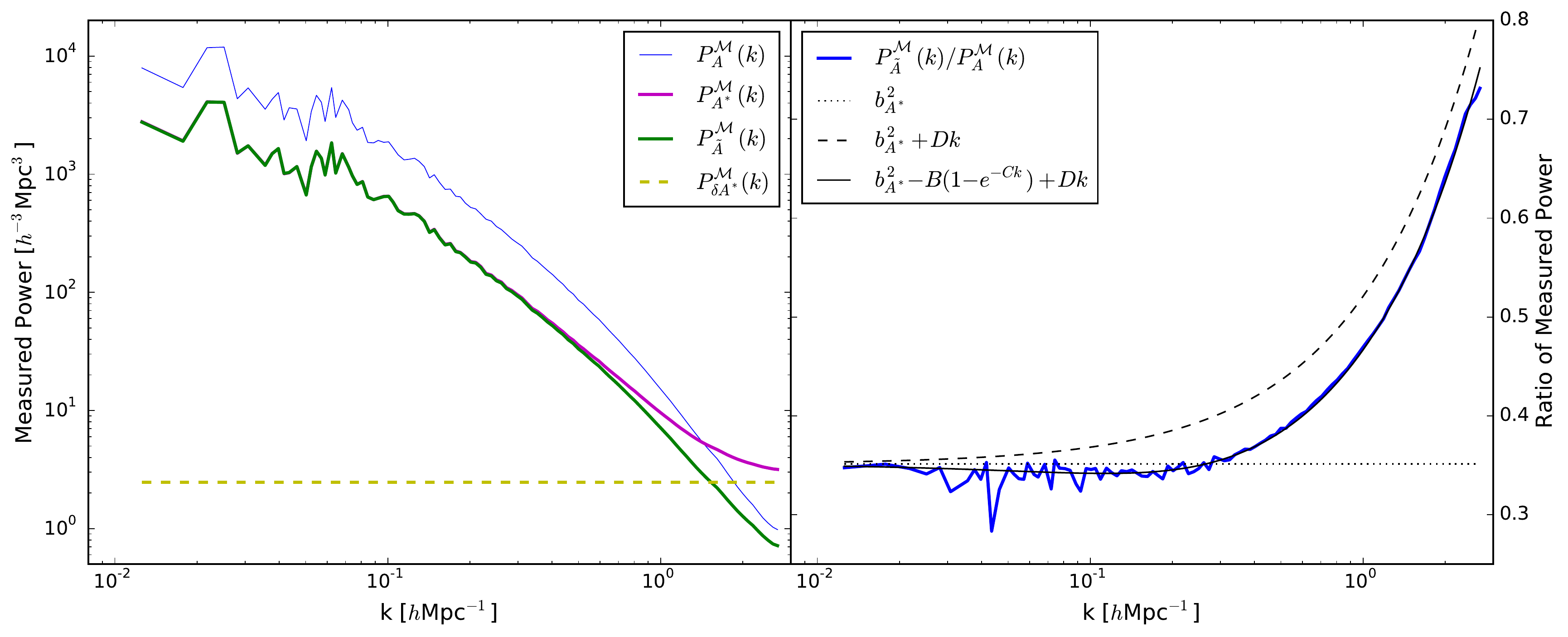}
\caption{Left panel: comparison of $P^\mathcal{M}_A(k)$ and $P^\mathcal{M}_{A^*}(k)$, along with the power spectra of the components of $A^*$, for a discrete realization of the Millennium Simulation. The $\delta\!A^*$ component encapsulates the stochasticity incurred by Poisson sampling and thus produces a discreteness plateau; the $\tilde{A}$ component encapsulates the nonlinearity of the $A \mapsto A^*$ map. Even after removing discreteness effects, the $\tilde{A}$-power spectrum exhibits a somewhat different shape than the $A$-power spectrum. (Note that these are the measured power spectra and thus include pixel-window and aliasing effects.) Right panel: the ratio of the measured $\tilde{A}$- and $A$-power spectra, compared to the bias only (dotted line), the bias plus a linear term (dashed line), and the bias plus a linear term and a decaying exponential (using best-fit values for $B$ and $D$ -- see Equation~\ref{eq:parametrization}). In both panels we use a Poisson sampling ($\overline{N}=1.0$) of the $z=0$ Millennium Simulation snapshot in the original Millennium Simulation cosmology.}
\label{fig:Atildecomponents}
\end{figure*}

At this point we introduce one subtlety that is of practical importance, namely, that the measured power spectrum will reflect the effects of pixelation and aliasing (see \citealp{Jing2005}). In the sequel we must explicitly distinguish between measured and theoretical spectra -- and thus we use $P^\mathcal{M}_A(k)$ and $P^\mathcal{M}_{\tilde{A}}(k)$ to denote the \emph{measured} power spectra (which include the pixel window and aliasing effects), and we retain the notation $P_A(k)$ and $P_{\tilde{A}}(k)$ for the theoretical spectra (such as those obtained from \textsc{Camb}), which do not include these effects. The relationships detailed in \citet{Jing2005} show how to account for these effects; in particular, it is fairly straightforward (numerically) to pass from $P_A(k)$ to $P^\mathcal{M}_A(k)$ (and likewise for $P^\mathcal{M}_{\tilde{A}}(k)$) -- see Equation~\ref{eq:P_waa}.

Since observations typically count galaxies instead of directly measuring dark matter, we will write our expression for $\tilde{A}$ in terms of the measured spectra (which include pixelation and aliasing effects); thus, our task is  to fit the difference between $P^\mathcal{M}_{\tilde{A}}(k) / P^\mathcal{M}_A(k)$ and the constant $b^2_{A^*}$ from Equations~\ref{eq:Atildebias} and \ref{eq:prelimPAs}. By comparing with discrete realizations of the Millennium Simulation (described in more detail later in this section), we observe that at small scales (large $k$) this ratio $P^\mathcal{M}_{\tilde{A}}(k) / P^\mathcal{M}_A(k)$ is virtually linear in $k$, and at large scales (small $k$) it matches a decaying exponential. We hence add two terms to $b^2_{A^*}$ and write
\begin{equation}
\frac{P^\mathcal{M}_{\tilde{A}}(k)}{P^\mathcal{M}_A(k)} = b^2_{A^*} - B\left(1 - e^{-Ck}\right) + Dk,
\label{eq:parametrization}
\end{equation}
where $B$, $C$, and $D$ are (possibly cosmology-dependent) factors to be determined (see Figure~\ref{fig:Atildecomponents}).

Considering the terms one by one, we see that at the largest scales (smallest $k$) the constant bias $b^2_{A^*}$ is dominant. The second term sets a scale $k \sim C^{-1}$ at which the ratio $P^\mathcal{M}_{\tilde{A}}(k)/P^\mathcal{M}_A(k)$ decreases to $b^2_{A^*} - B$. The final term indicates that at small scales (large $k$) the passage from $A$ to $\tilde{A}$ produces more power than a simple bias would produce, and $D$ (with units of length) parametrizes this increase. Our task is somewhat eased by the fact that the fit is not extremely sensitive to the values of any one of these parameters: experimentation shows enough degeneracy among them that changes in the value of one parameter can often be offset by a change in the value of another, without substantially affecting the overall accuracy of the fit.

To obtain reasonable values for these parameters, we note from numerical experiments that the quality of the fit is not particularly sensitive to the value of $C$. Thus, based on typical best-fit values (when fitting all three parameters simultaneously), we note that we can employ a constant value for
\begin{equation}
C^{-1} = 0.15h \mbox{ Mpc}^{-1}
\label{eq:C}
\end{equation}
to set the scale for the onset of the shape change caused by the second term of Equation~\ref{eq:parametrization}. Experimentation with other values of $C$ -- including allowing for a redshift-dependence -- did not seem to materially affect the accuracy of our fits. This insensitivity is presumably due to the fact that $B$ and $C$ are somewhat degenerate, and, as detailed later, we calculate $B$ from a given value of $C$ (and of other parameters).

\begin{figure}
\leavevmode\epsfxsize=9cm\epsfbox{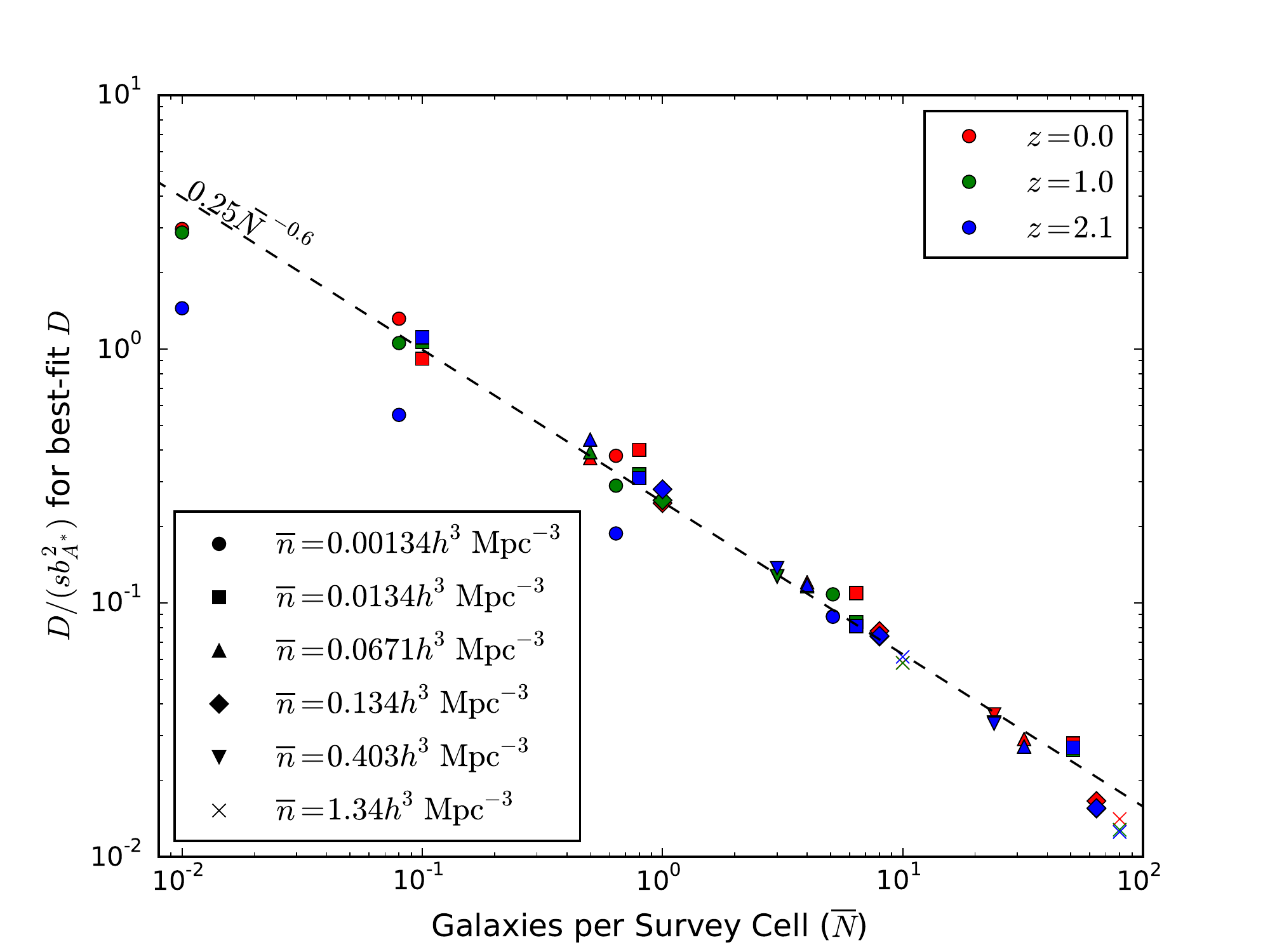}
\caption{Best-fit values of the $D$-parameter (see Equation~\ref{eq:parametrization}) for $P^\mathcal{M}_{\tilde{A}}(k)$, normalized by survey cell side length $s$ and $A^*$-bias $b_{A^*}^2$, for a variety of redshifts, number densities, and smoothing scales. The dashed line shows the fitting formula adopted in Equation~\ref{eq:D}. Data points are from discrete realizations of the Millennium Simulation.}
\label{fig:D_plot}
\end{figure}

We next characterize the parameter $D$. To do so, we obtain Millennium Simulation \citep{MillSim} snapshots\footnote{http://gavo.mpa-garching.mpg.de/Millennium/} from $z = 0.0$, 1.0, and 2.1; these snapshots utilize cubical survey cells of side length $500h^{-1}\mathrm{Mpc}/256\,$cells $= 1.95h^{-1}$Mpc/cell. We then create discrete realizations (via Poisson sampling) at number densities $\overline{N} = 0.01,$ 0.1, 0.5, 1.0, 3.0, and 10.0 galaxies per cell. We also smooth these realizations by binning them on scales ranging from the original $1.95h^{-1}$Mpc/cell up to $31.25h^{-1}$Mpc/cell. After calculating the power spectrum $P^\mathcal{M}_{\tilde{A}}(k)$ for these realizations, we fit the power spectrum with Equation~\ref{eq:parametrization} (fixing  $C$ to the value in Equation~\ref{eq:C}). We thus obtain a set of best-fit values (using least-squares optimization) for the parameters $B$ and $D$. It is these best-fit values for $D$ which appear in Figure~\ref{fig:D_plot}.

The parameter $D$ has units of length, and the most obvious length scale is the pixel size $s$. It is also reasonable to suppose that $D$ depends on the constant bias term $b^2_{A^*}$. We note that if we normalize $D$ by $s b^2_{A^*}$, we obtain a power-law relationship between $D/sb^2_{A^*}$ and $\overline{N}$ (the average number of galaxies per cell, after smoothing) shown in Figure~\ref{fig:D_plot}. We thus propose the following fitting formula for $D$:
\begin{equation}
\label{eq:D}
D = \frac{s \, b^2_{A^*}}{4\overline{N}^{\,0.6}},
\end{equation}
where $s$ is the side length of the (cubical) survey cell and $\overline{N}$ is the average number of galaxies per cell. (This relationship appears as a dashed line on Figure~\ref{fig:D_plot}.) We note that for extremely low number densities and high redshifts, Equation~\ref{eq:D} appears to overpredict the best-fit value of $D$; however, the practical utility of the power spectrum is limited to cases in which $\overline{N} \ga 1.$ In any case, the rationale for the additional factor and exponent in Equation~\ref{eq:D} is purely pragmatic, to be justified by whether or not it ultimately produces a reasonable fit to the $A^*$-power spectra of the Millennium Simulation data (including the rescalings we consider in Section~\ref{sec:accuracy}).

\begin{figure}
\leavevmode\epsfxsize=9cm\epsfbox{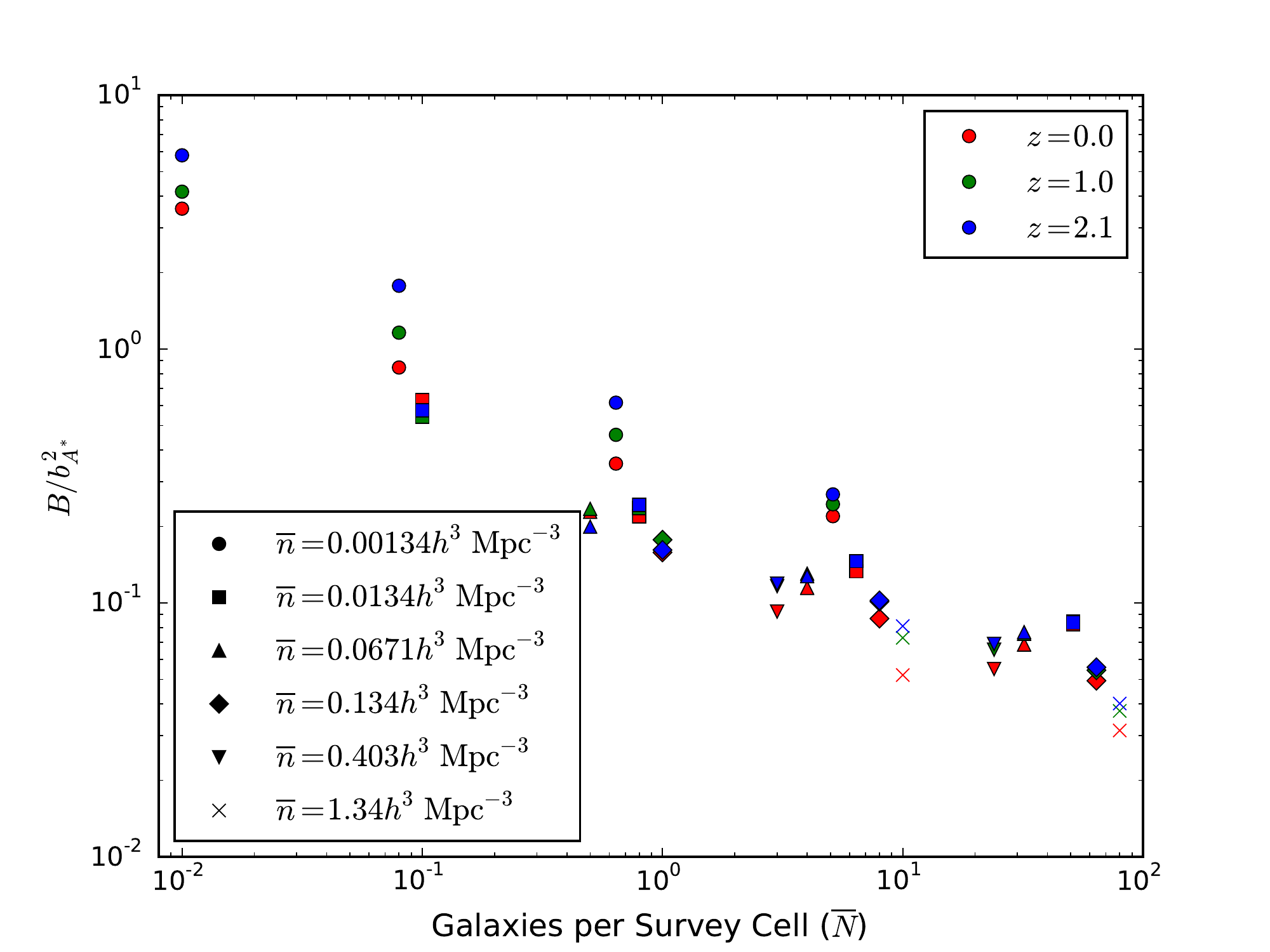}
\caption{Values of the $B$-parameter (see Equation~\ref{eq:parametrization}) for $P^\mathcal{M}_{\tilde{A}}(k)$, normalized by the $A^*$-bias $b_{A^*}^2$, computed using Equations~\ref{eq:D} and \ref{eq:Bbetter}. Data points are calculated from discrete realizations of the Millennium Simulation.}
\label{fig:B_plot}
\end{figure}

Figure~\ref{fig:B_plot} indicates that the best-fit values of the $B$-parameter also (roughly) follow a power law in $\overline{N}$.
However, now that we have an analytic approximation for $D$, we need not derive an analogous expression for $B$; rather, we can utilize the fact that the integral of the power spectrum yields the variance. Thus, if $V_k$ is the survey volume in Fourier space, the relationship
\begin{equation}
\begin{split}
\sigma^2_{\tilde{A}} & = \int_V \frac{d^3k}{(2\pi)^3} P^\mathcal{M}_{\tilde{A}}(k) \\
	& = \int_{V_k} \frac{d^3k}{(2\pi)^3} \left( b^2_{A^*} - B\left(1 - e^{-Ck}\right) + Dk\right) P^\mathcal{M}_A(k)
\end{split}
\label{eq:Bbetter}
\end{equation}
(together with Equations~\ref{eq:sig2Atilde}, \ref{eq:C}, and \ref{eq:D}) permits calculation of $B$.

Using Equation~\ref{eq:Bbetter} to calculate $B$ (after using Equations~\ref{eq:C} and \ref{eq:D} to determine $C$ and $D$), we obtain the values displayed in Figure~\ref{fig:B_plot}. Note that the figure does not display the $B$-values obtained by simultaneously fitting $D$ and $B$, but rather the $B$-values necessary to obtain the correct variance given our approximation for $D$.

\subsection{A Recipe for Calculating $P^\mathcal{M}_{A^*}(k)$}
\label{sec:recipe}

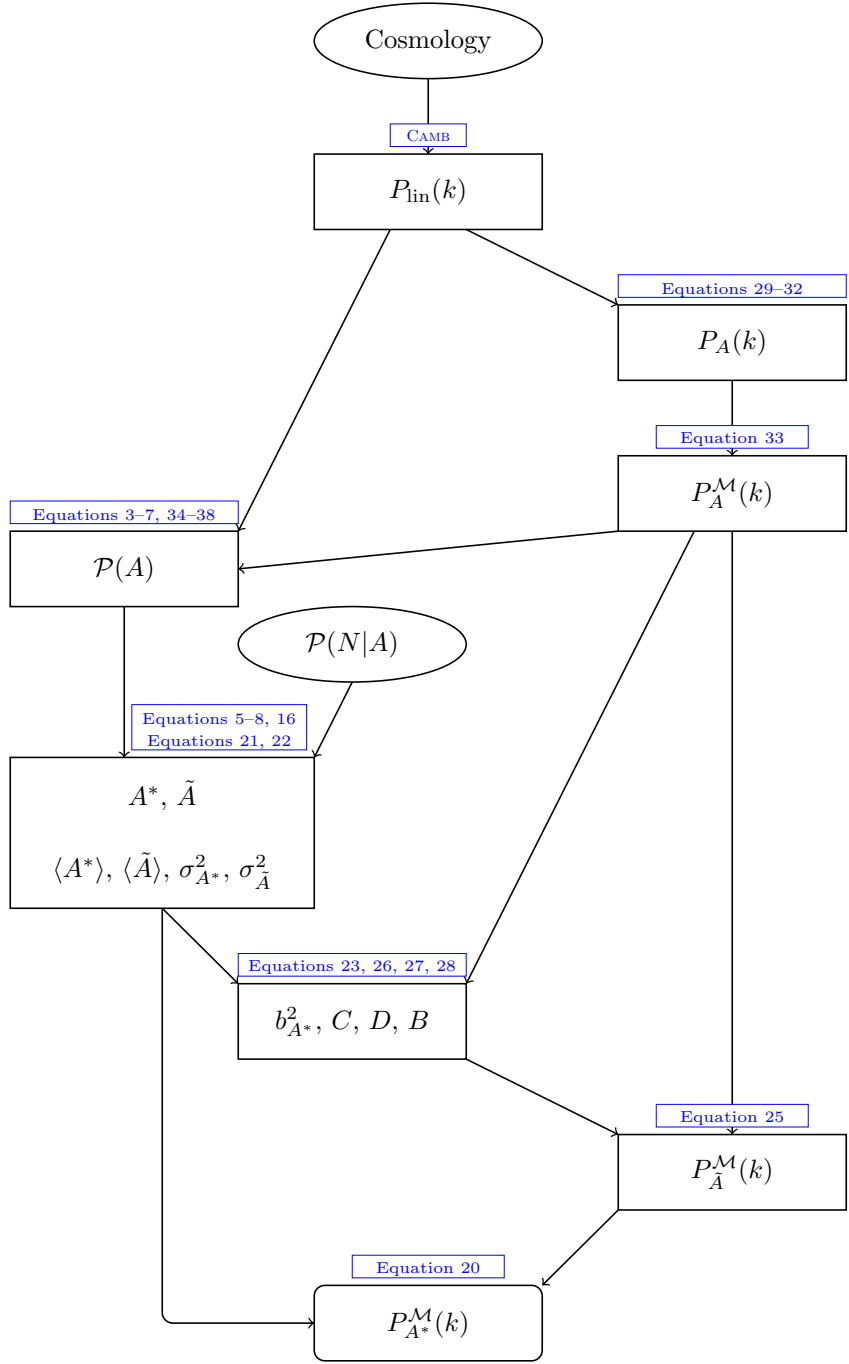
\begin{figure*}
\leavevmode\epsfxsize=18cm
\begin{tikzpicture}

\draw[semithick] (5.5,17.5) ellipse (1.5 and 0.5);
\node[font=\large] at (5.5,17.5) {Cosmology};

\draw[semithick, ->] (5.5,17) -- (5.5,16);

\draw[ultra thin, blue, fill=white] (5,16.1) rectangle (6,16.4);
\node[font=\scriptsize, blue] at (5.5,16.25) {\textsc {Camb}};
\draw[semithick] (4,15) rectangle (7,16);
\node[font=\large] at (5.5,15.5) {$P_\mathrm{lin}(k)$};

\draw[semithick, ->] (5,15) -- (3,11);
\draw[semithick, ->] (6,15) -- (8,14);

\draw[ultra thin, blue, fill=white] (8,14.1) rectangle (11,14.4);
\node[font=\scriptsize, blue] at (9.5,14.2) {Equations \ref{eq:PAk}--\ref{eq:N}};
\draw[semithick] (8,13) rectangle (11,14);
\node[font=\large] at (9.5,13.5) {$P_A(k)$};


\draw[ultra thin, blue, fill=white] (0,11.1) rectangle (3,11.4);
\node[font=\scriptsize, blue] at (1.5,11.2) {Equations \ref{eq:GEV}--\ref{eq:muG}, \ref{eq:sigA}--\ref{eq:pns}};
\draw[semithick] (0,10) rectangle (3,11);
\node[font=\large] at (1.5,10.5) {$\mathcal{P}(A)$};

\draw[semithick, ->] (9.5,13) -- (9.5,12);

\draw[ultra thin, blue, fill=white] (8.5,12.1) rectangle (10.5,12.4);
\node[font=\scriptsize, blue] at (9.5,12.22) {Equation \ref{eq:P_waa}};
\draw[semithick] (8,11) rectangle (11,12);
\node[font=\large] at (9.5,11.5) {$P_A^\mathcal{M}(k)$};

\draw[semithick, ->] (8,11) -- (3,10.5);

\draw[semithick] (4.5,9.5) ellipse (1.5 and 0.5);
\node[font=\large] at (4.5,9.5) {$\mathcal{P}(N|A)$};

\draw[semithick, ->] (1.5,10) -- (1.5,8);
\draw[semithick, ->] (4.5,9) -- (4,8);

\draw[ultra thin, blue, fill=white] (1.6,8.1) rectangle (3.9,8.7);
\node[font=\scriptsize, blue] at (2.75,8.5) {Equations \ref{eq:xiG}--\ref{eq:AstarGEV}, \ref{eq:Atilde}};
\node[font=\scriptsize, blue] at (2.75,8.2) {Equations \ref{eq:sig2Astar}, \ref{eq:sig2Atilde}};
\draw[semithick] (0,6) rectangle (4,8);
\node[font=\large] at (2,7.5) {$A^*$, $\tilde{A}$};
\node[font=\large] at (2,6.5) {$\langle A^* \rangle$, $\langle \tilde{A} \rangle$, $\sigma_{A^*}^2$, $\sigma_{\tilde{A}}^2$};

\draw[semithick, ->] (2,6) -- (3,5);
\draw[semithick, ->] (9,11) -- (6,5);

\draw[ultra thin, blue, fill=white] (3,5.1) rectangle (6,5.4);
\node[font=\scriptsize, blue] at (4.5,5.22) {Equations \ref{eq:Atildebias}, \ref{eq:C}, \ref{eq:D}, \ref{eq:Bbetter}};
\draw[semithick] (3,4) rectangle (6,5);
\node[font=\large] at (4.5,4.5) {$b_{A^*}^2$, $C$, $D$, $B$};

\draw[semithick, ->] (9.5,11) -- (9.5,3);
\draw[semithick, ->] (6,4) -- (8,3);

\draw[ultra thin, blue, fill=white] (8.5,3.1) rectangle (10.5,3.4);
\node[font=\scriptsize, blue] at (9.5,3.22) {Equation \ref{eq:parametrization}};
\draw[semithick] (8,2) rectangle (11,3);
\node[font=\large] at (9.5,2.5) {$P_{\tilde{A}}^\mathcal{M}(k)$};

\draw[semithick, ->] (8,2) -- (7,1);
\draw[semithick, rounded corners, ->] (2,6) -- (2,0.5) -- (4,0.5);

\draw[ultra thin, blue, fill=white] (4.5,1.1) rectangle (6.5,1.4);
\node[font=\scriptsize, blue] at (5.5,1.22) {Equation \ref{eq:PAswithPdeltAs}};
\draw[semithick, rounded corners] (4,0) rectangle (7,1);
\node[font=\large] at (5.5,0.5) {$P_{A^*}^\mathcal{M}(k)$};

\end{tikzpicture}
\caption{Block diagram of procedure for calculating the optimal galaxy power spectrum $P_{A^*}(k)$. The required inputs (besides survey parameters such as redshift) are the cosmology and the discretization scheme $\mathcal{P}(N|A)$.}
\label{fig:block_diag}
\end{figure*}

We now have all the information necessary to calculate $P^\mathcal{M}_{A^*}(k)$. We summarize the process below and present the same information schematically in Figure~\ref{fig:block_diag}. For ease of reference, we here reproduce the relevant equations with their original equation numbers.

The required inputs for the process (besides survey parameters such as redshift) are (1) the underlying cosmology and (2) the discrete sampling scheme $\mathcal{P}(N|A)$, which is the probability distribution for galaxy counts given an underlying value of $A$.

The first step is to obtain the linear power spectrum using \textsc{Camb} or similar software. From $P_\mathrm{lin}(k)$ one can then derive the log spectrum $P_A(k)$ using the following prescription of \citet{ReppPAk}:
\begin{equation}
\label{eq:PAk}
P_A(k) = NC_\mathrm{corr}(k) \cdot \frac{\mu}{\sigma_\mathrm{lin}^2} \ln \left(1+\frac{\sigma_\mathrm{lin}^2}{\mu}\right) \cdot P_\mathrm{lin}(k),
\end{equation}
with the best-fit value $\mu=0.73$, and where one calculates the linear variance by
\begin{equation}
\label{eq:sig2lin}
\sigma_\mathrm{lin}^2 = \int_0^{k_N} \frac{dk\,k^2}{2\pi^2} P_\mathrm{lin}(k).
\end{equation}
In this equation, $k_N$ is the Nyquist frequency $\pi/\ell$, where $\ell$ is the side length of one pixel of the survey volume. $C_\mathrm{corr}(k)$ in Equation~\ref{eq:PAk} is a slope correction\footnote{Unrelated to the parameter $C$ from Equations~\ref{eq:parametrization} and \ref{eq:C}} with normalization $N$, both of which are given by the following equations:
\begin{equation}
\label{eq:slopemod}
C_\mathrm{corr}(k)=\left\{
\begin{array}{ll}
    1	&	\mbox{if $k < 0.15h$ Mpc$^{-1}$} \\
    (k/0.15)^\alpha	& \mbox{if $k \ge 0.15h$ Mpc$^{-1}$}
\end{array}
\right.,
\end{equation}
\begin{equation}
\label{eq:N}
N=\frac{\int dk\, k^2 P_\mathrm{lin}(k)}{\int dk\, k^2 C_\mathrm{corr}(k) P_\mathrm{lin}(k)}.
\end{equation}
Appropriate values of $\alpha$ range from 0.02 at $z=0$ to 0.14 at $z=2.1$ (see table~1 of \citealp{ReppPAk}).

Next, the prescription of \citet{Jing2005} allows us to obtain the ``measurable'' $P^{\mathcal{M}}_A(k)$, which includes pixel window and alias effects:
\begin{equation}
P^\mathcal{M}_{A}(k) = \left\langle\sum_{\mathbf{n} \in \mathbb{Z}^3} P_A(\mathbf{k}+2k_N\mathbf{n}) W(\mathbf{k}+2k_N\mathbf{n})^2\right\rangle_{|\mathbf{k}|=k};
\label{eq:P_waa}
\end{equation}
here the sum runs over all three-dimensional integer vectors $\mathbf{n}$, though we find it sufficient to consider only $|\mathbf{n}| < 3$.

At this point we can obtain the moments of the log probability distribution -- and the distribution itself -- by using the GEV prescription of \citet{ReppApdf}. First, the variance:
\begin{equation}
\sigma^2_A = \int_{V_k\setminus\{0\}} \frac{d^3k}{(2\pi)^3} P^\mathcal{M}_{A}(k),
\label{eq:sigA}
\end{equation}
where the region denoted $V_k\setminus\{0\}$ is the set of non-zero $\mathrm{k}$-vectors corresponding to the real-space volume of the survey. Next, the mean:
\begin{equation}
\langle A \rangle = -\lambda \ln \left(1+\frac{\sigma^2_\mathrm{lin}(k_N)}{2\lambda}\right),
\label{eq:meanAfit}
\end{equation}
where the best-fit value of $\lambda$ is 0.65. Finally, the skewness:
\begin{equation}
\gamma_1 \equiv \frac{\langle\left(A-\langle A \rangle\right)^3\rangle}{\sigma_A^3}
\end{equation}
\begin{equation}
\gamma_1 = \left(a(n_s+3) + b\right) \left( \sigma_A^2 \right)^{-p(n_s) + 1/2}
\end{equation}
\begin{equation}
p(n_s) = d + c \ln(n_s+3)\label{eq:pns},
\end{equation}
where $n_s$ is the slope of the linear no-wiggle power spectrum of \citet{EisensteinHu}, and the best values of the parameters are $a=-0.70$, $b=1.25$, $c=-0.26$, and $d=0.06$.

One can now calculate the log probability distribution $\mathcal{P}(A)$, as explained previously:
\begin{equation}
\tag{\ref{eq:GEV}}
\mathcal{P}(A) = \frac{1}{\sigma_G} t(A)^{1+\xi_G} e^{-t(A)}
\end{equation}
\begin{equation}
\tag{\ref{eq:GEV_t}}
t(A) = \left(1 + \frac{A - \mu_G}{\sigma_G}\xi_G\right)^{-1/\xi_G}
\end{equation}
\begin{equation}
\gamma_1 = -\frac{\Gamma(1-3\xi_G) - 3\Gamma(1-\xi_G)\Gamma(1-2\xi_G) + 2\Gamma^3(1-\xi_G)}{\left(\Gamma(1-2\xi_G) - \Gamma^2(1-\xi_G)\right)^{3/2}}
\tag{\ref{eq:xiG}}
\end{equation}
\begin{equation}
\sigma_G = \sigma_A \xi_G \cdot \left(\Gamma(1-2\xi_G) - \Gamma^2(1-\xi_G)\right)^{-1/2}
\tag{\ref{eq:sigG}}
\end{equation}
\begin{equation}
\mu_G = \langle A \rangle - \sigma_G \frac{\Gamma(1-\xi_G) - 1}{\xi_G}.
\tag{\ref{eq:muG}}
\end{equation}

The next step is to calculate the first two moments of $A^*$ and $\tilde{A}$, which in turn require expressions for these two quantities. The relevant equations (assuming a GEV log matter distribution) are as follows:
\begin{multline}
\tag{\ref{eq:AstarGEV}}
\frac{1}{\sigma_G} \left( 1 + \frac{A^*(N) - \mu_G}{\sigma_G} \xi_G \right)^{-1-\frac{1}{\xi_G}} + N \\
= \frac{1+\xi_G}{\sigma_G + \left(A^*(N) - \mu_G\right)\xi_G} + \overline{N}e^{A^*(N)}
\end{multline}
\begin{equation}
\tilde{A}(A) = \sum_N \mathcal{P}(N|A) A^*(N)
\tag{\ref{eq:Atilde}}
\end{equation}
\begin{equation}
\sigma^2_{A^*} = \sum_N \int dA\,\,\mathcal{P}(A) \mathcal{P}(N|A)\left(A^*(N) - \langle A^* \rangle \right)^2 \tag{\ref{eq:sig2Astar}}
\end{equation}
\begin{equation}
\sigma^2_{\tilde{A}} = \int dA\,\,\mathcal{P}(A) \left(\tilde{A}(A) - \langle \tilde{A} \rangle \right)^2 \tag{\ref{eq:sig2Atilde}}.
\end{equation}
Recall that $\xi_G$, $\sigma_G$, and $\mu_G$ are the parameters of the distribution $\mathcal{P}(A)$, related to the moments of $A$ by Equations~\ref{eq:xiG}--\ref{eq:muG}. The first moments $\langle A^* \rangle$ and $\langle \tilde{A} \rangle$ are calculated using integrals analogous to those in Equations~\ref{eq:sig2Atilde} and \ref{eq:sig2Astar}.

From these moments and from $P^{\mathcal{M}}_A(k)$, one can then obtain the parameters $b_{A^*}^2$, $C$, $D$, and $B$ for $P^\mathcal{M}_{\tilde{A}}(k)$:
\begin{equation}
b^2_{A^*} = \frac{1}{\sigma_A^4} \left\lbrace \int dA\,(A-\overline{A})\left(\tilde{A}(A) - \langle\tilde{A}(A) \rangle\right)\mathcal{P}(A) \right\rbrace^2
\tag{\ref{eq:Atildebias}}
\end{equation}
\begin{equation}
C^{-1} = (1/0.15)h^{-1} \mbox{Mpc}
\tag{\ref{eq:C}}
\end{equation}
\begin{equation}
\tag{\ref{eq:D}}
D = \frac{s \, b^2_{A^*}}{4\overline{N}^{\,0.6}}
\end{equation}
\begin{equation}
\tag{\ref{eq:Bbetter}}
\sigma^2_{\tilde{A}} = \int_{V_k} \frac{d^3k}{(2\pi)^3}\left( b^2_{A^*} - B\left(1 - e^{-Ck}\right) + Dk\right) P^\mathcal{M}_A(k)
\end{equation}

These four parameters -- along with $P^{\mathcal{M}}_A(k)$ -- then yield the spectrum of $\tilde{A}$:
\begin{equation}
P^\mathcal{M}_{\tilde{A}}(k) = \left[b^2_{A^*} - B\left(1 - e^{-Ck}\right) + Dk\right] \cdot P^\mathcal{M}_A(k).
\tag{\ref{eq:parametrization}}
\end{equation}

Finally, one must add the discreteness plateau to obtain the power spectrum of $A^*$ itself:
\begin{equation}
P^\mathcal{M}_{A^*}(k) = P^\mathcal{M}_{\tilde{A}}(k) + \delta V \cdot \left(\sigma^2_{A^*} - \sigma^2_{\tilde{A}}\right).
\tag{\ref{eq:PAswithPdeltAs}}
\end{equation}

\section{Accuracy}
\label{sec:accuracy}
It remains to evaluate the accuracy of the prescription in Sections~\ref{sec:bias}--\ref{sec:shape}.

\begin{figure*}
\leavevmode\epsfxsize=18cm\epsfbox{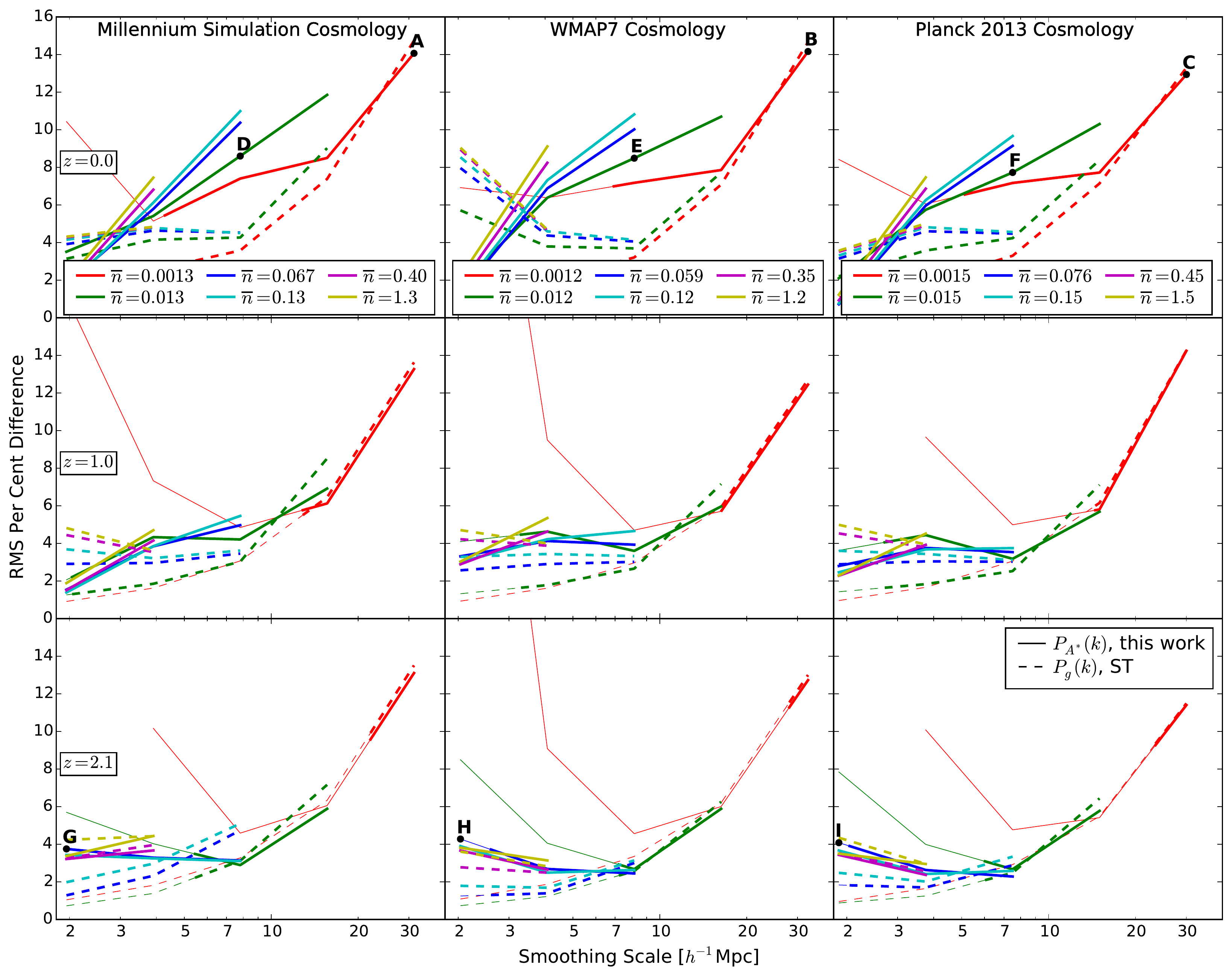}
\caption{Solid lines show RMS per cent differences (weighted by cosmic variance) between our prescription for $P^\mathcal{M}_{A^*}(k)$ (Section~\ref{sec:shape}) and the values of $P^\mathcal{M}_{A^*}(k)$ measured from discrete realizations of the Millennium Simulation. We show results for three near-concordance cosmologies (columns) and three redshifts (rows), for a variety of scales (side lengths of cubical pixels -- shown on the horizontal axes) and number densities ($\overline{n}$, in units of $h^3$Mpc$^{-3}$). For the lowest number densities, the transition from thick to thin lines marks the Poisson noise limit -- the scale at which $P(k) = 1/\overline{n}$. Dashed lines show the corresponding results for the ST (\citealt{Smith_et_al}/\citealt{Takahashi2012}) prescription for $P^\mathcal{M}_g(k)=P^\mathcal{M}(k) + 1/\overline{n}$ (as implemented in \textsc{Camb}), again as compared to values measured in the Millennium Simulation. Note that the high ``error'' apparent at large scales is almost wholly due to the dominance of cosmic variance on these scales. Capital letters denote spectra displayed in the corresponding panels of Figure~\ref{fig:SelectSpec}.}
\label{fig:RMSErrors}
\end{figure*}

\begin{figure*}
\leavevmode\epsfxsize=18cm\epsfbox{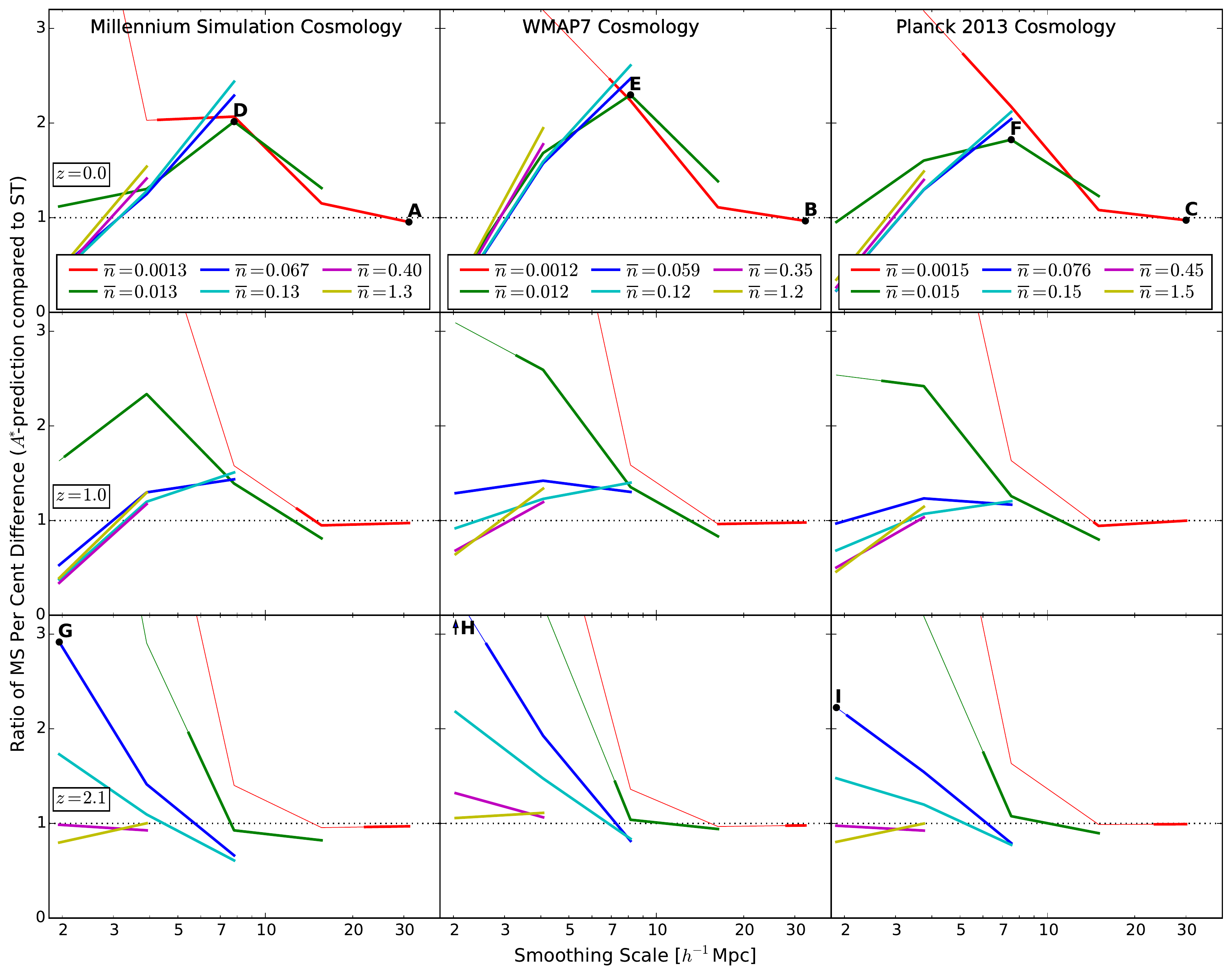}
\caption{Ratios between the two sets of weighted RMS per cent-difference calculations in Figure~\ref{fig:RMSErrors}. In particular, each data point shows the ratio of (1) the (weighted) RMS per cent difference between our prescription (for $P^\mathcal{M}_{A^*}(k)$) and the Millennium Simulation, and of (2) the prescription of ST (\citealt{Smith_et_al}/\citealt{Takahashi2012} -- for $P^\mathcal{M}_g(k)$) and the Millennium Simulation. Columns and rows are as in Figure~\ref{fig:RMSErrors}; transitions from thick to thin lines denote Poisson noise limits; and capital letters denote spectra displayed in the corresponding panels of Figure~\ref{fig:SelectSpec}.}
\label{fig:RMSError_ratios}
\end{figure*}

To do so, we obtain (as described previously in Section~\ref{sec:shape}) snapshots of the Millennium Simulation at $z=0$, 1.0, and 2.1. These snapshots comprise $256^3$ cubical pixels with side lengths $1.95h^{-1}$Mpc. We then Poisson-sample the dark matter density to obtain discrete realizations of each snapshot for mean number of particles per pixel $\overline{N} = 0.01, 0.1, 0.5, 1.0, 3.0$, and 10.0. For $\overline{N} \ge 0.5$, we generate 10 realizations per redshift; for $\overline{N} = 0.1$ and 0.01, we generate 20. This ensures an overall sampling variance (in each pixel) of at most $0.5(1+\delta)^{-1}$, except for $\overline{N}=0.01$. However, the $\overline{N}=0.01$ case is of little practical importance until we rebin the realizations (see below) on scales of $\sim 4h^{-1}$Mpc (see also extent of thick line in  Figures~\ref{fig:PkdeltAstar} and \ref{fig:RMSErrors}), at which point the pixel variance reaches $0.6(1+\delta)^{-1}$, comparable to the other number densities.

Since the $A^*$-power spectrum is scale-dependent, we must investigate multiple smoothing scales. To do so, we take each of our discrete realizations and rebin it to two, four, eight, and sixteen times the original pixel length, reaching a maximum scale of $31.25h^{-1}$Mpc. For each of these rebinnings, we calculate $P^\mathcal{M}_{A^*}(k)$, provided that the mean number of galaxies per cell $\overline{N} < 100$; at higher number densities, $A^*$ differs little from $A$, and the computation of the $A^*$-moments becomes expensive.

The procedure so far allows us to test our prescription for only the original Millennium Simulation cosmology. However, \citet{AnguloWhite} outline a method for re-scaling simulations from one cosmology to another by matching linear variances; doing so involves both a re-scaling of survey cell size and a re-mapping of simulation snapshots to redshift. Such rescalings of the Millennium Simulation to the WMAP7 and Planck 2013 cosmologies are publicly available. Therefore we repeat the above procedure ($z=0.0,1.0,2.1$, for $\overline{N}$ at the original pixel scale from 0.01 to 10.0, rebinned to scales from 1 to 16 times the original pixel length) for both of these rescalings. Hence we end up with simulations in three redshifts, with multiple number densities, at scales ranging from $2h^{-1}$ to $32h^{-1}$Mpc, and for three near-concordance cosmologies.

These discrete realizations provide us with a standard against which to compare our prescription. We implement our prescription using \textsc{Camb} (Code for Anisotropies in the
Microwave Background:\footnote{http://camb.info/} \citealp{CAMB}) to generate the appropriate linear power spectra $P_\mathrm{lin}(k)$, from which we obtain $P_A(k)$ by following the prescription presented in \citet{ReppPAk}. We can then use the work summarized at the end of Section~\ref{sec:shape} to predict the power spectrum $P^\mathcal{M}_{A^*}(k)$.

We also wish to compare the accuracy of our prescription to that of \citet{Smith_et_al}/\citet{Takahashi2012} (hereafter ST), which is the prescription used in \textsc{Camb} for nonlinear spectra. To do so, we measure the galaxy power spectra $P^\mathcal{M}_g(k)$ of the various realizations; and we obtain the ST prescription by using \textsc{Camb} to calculate $P_g(k) = P(k) + 1/\overline{n}$ and then following the method of \citet{Jing2005} to get the predicted $P^\mathcal{M}_g(k)$.

Our metric for comparison is the root-mean-square (RMS) per cent difference between the predicted and measured power spectra (using logarithmically-spaced $k$-values). To mitigate the effect of the increase in cosmic variance at large scale -- and the resultant power spectrum stochasticity -- due to the limited number of $k$-modes included in the Millennium Simulation volume at such scales, we weight the mean by the inverse cosmic  variance at each $k$-value. Thus,
\begin{equation}
\mathrm{RMS} = \sqrt{ \sum_k \frac{\left( M(k) - T(k) \right)^2 }{\sigma_{\mathrm{CV}}^2(k)} \left/ \sum_k \frac{1}{\sigma_{\mathrm{CV}}^2(k)} \right. },
\end{equation}
where $M(k)$ is the power spectrum value measured from our realizations, $T(k)$ is the value predicted by our recipe, and
$\sigma_{\mathrm{CV}}(k) = P_{(A^*\mbox{\textrm{ \scriptsize{or}} }g)}(k)/\sqrt{N_k}$ is the cosmic variance (or technically, standard deviation) at a given $k$-mode for a given power spectrum value ($P_{A^*}(k)$ or $P_g(k)$) determined from a given number $N_k$ of modes. We take the sum over logarithmically-spaced $k$-values: the simulation size sets $k_{\mathrm{min}}=2\pi/\ell$, where $\ell$ is the length of one side of the simulation cube ($500h^{-1}$ Mpc for the original Millennium Simulation); the pixel size sets $k_{\mathrm{max}}=k_N\sqrt{3}=\pi\sqrt{3}/s$, where $s$ is the smoothing scale (i.e., the side length of a cubical pixel) -- this being the largest $k$ measurable from such pixels. The value of $\sigma_{\mathrm{CV}}$ represents only the cosmic variance inherent in the Millennium Simulation, not the variance from the Poisson sampling process (which we reduce by averaging multiple Poisson realizations). Note that even with this weighting, cosmic variance will dominate the calculated RMS ``error'' for large smoothing scales.

The results appear in Figure~\ref{fig:RMSErrors}. In this plot, we again use a transition from thick to thin lines to indicate the Poisson shot noise limit, at which $P(k)$ becomes less than $1/\overline{n}$; we restrict our focus to scales above this limit. We find that for small smoothing scales ($\la 4h^{-1}$ Mpc) our recipe performs quite well, with accuracy to a few per cent, comparable to that of ST. At large scales ($\ga 15h^{-1}$ Mpc) the per cent difference (we should not in this case call it ``error'') increases greatly because of cosmic variance, both for our prescription and for that of ST; nevertheless, our accuracy is comparable to that of ST at these scales. At intermediate scales ($\sim 8h^{-1}$ Mpc) at low redshifts we find, however, a pronounced increase in the per cent difference with respect to ST.

One way to account for the large-scale cosmic variance is to divide each of the per cent differences from our recipe by the corresponding per cent difference from ST, thus obtaining a ratio of the prescriptions' accuracies. These ratios appear in Figure~\ref{fig:RMSError_ratios} and confirm that the worst accuracy of our prescription comes from smoothing on scales $\sim 8h^{-1}$ Mpc, whereas at larger scales the accuracy is virtually indistinguishable from that of ST. At the smallest scales, the accuracy is typically better than ST except at higher redshifts. However, reference to Figure~\ref{fig:RMSErrors} shows that even in these cases, the error in our prescription is still only a few per cent.

Thus in general, we find that the RMS error of our prescription is comparable to that of ST, or on the level of a few per cent. The exception would be at low redshifts ($z \sim 0$) at scales on the order of $8h^{-1}$ Mpc. Even in these cases, the error is typically less than 10 per cent. Furthermore, the low survey volume available at these redshifts gives them less weight in a survey designed for precise constraint of cosmological parameters.

\begin{figure*}
\leavevmode\epsfxsize=18cm\epsfbox{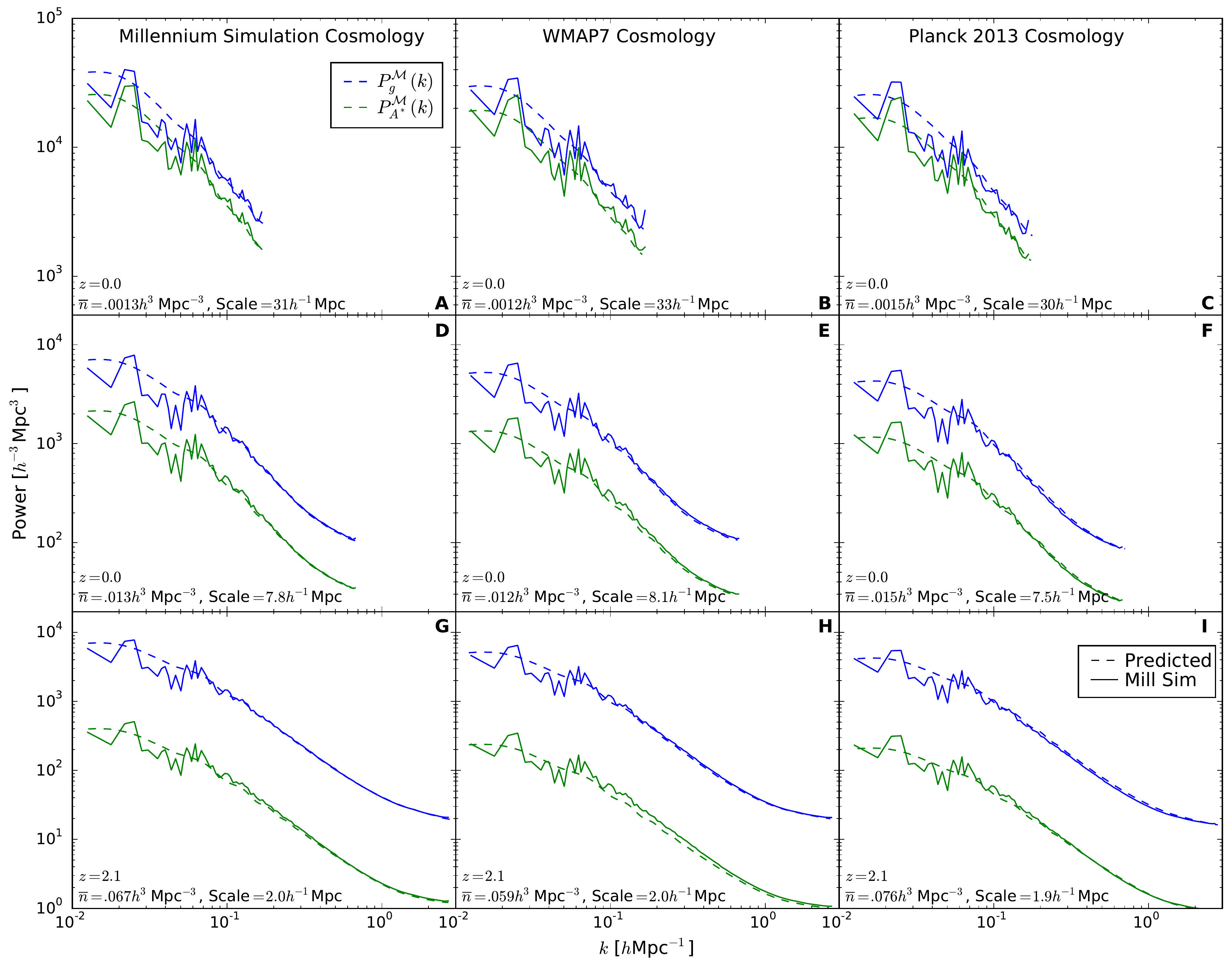}
\caption{Predicted versus simulated values for select power spectra -- for the Millennium, WMAP7, and Planck 2013 cosmologies (left, center, and right columns, respectively), for $z=0$ (top and middle rows) and 2.1 (bottom row), for the indicated number densities and smoothing scales. Blue curves show the galaxy overdensity power spectra $P^\mathcal{M}_g(k)$ as predicted by ST (dashed) and as measured from discrete Millennium Simulation realizations (solid); green curves show $P^\mathcal{M}_{A^*}(k)$ as calculated in Section~\ref{sec:shape} of this work (dashed) and as measured from the same discrete Millennium Simulation realizations (solid). Capital letters refer to the corresponding errors and error ratios labeled in Figures~\ref{fig:RMSErrors} and \ref{fig:RMSError_ratios}.}
\label{fig:SelectSpec}
\end{figure*}

We finally focus on nine specific examples of interest (denoted by letters A through I on Figures~\ref{fig:RMSErrors} and \ref{fig:RMSError_ratios}); we display the $A^*$- and galaxy-power spectra for these examples in Figure~\ref{fig:SelectSpec}.

The first three spectra (letters A--C in Figures~\ref{fig:RMSErrors}--\ref{fig:SelectSpec}, and the top row of panels in Figure~\ref{fig:SelectSpec}) correspond to number densities roughly equivalent to those anticipated for \textit{Euclid} and WFIRST, smoothed on scales $\sim 30h^{-1}$ Mpc. At these scales and densities, the accuracy of our prescription is virtually identical to that of ST. Indeed, reference to Figure~\ref{fig:RMSError_ratios} demonstrates that at $z \ga 1$, the accuracy of our prescription for such a Euclid-like survey is virtually indistinguishable from that of ST, given the impact of Poisson noise.

The next three spectra (letters D--F in Figures~\ref{fig:RMSErrors}--\ref{fig:SelectSpec}, and the middle row of panels in Figure~\ref{fig:SelectSpec}) investigate the anomalously high per cent differences encountered at intermediate smoothing scales and low redshifts. It appears that this difference is the result of higher-than-predicted power around $k \sim 0.1$--$0.2h$ Mpc$^{-1}$, although a significant amount of cosmic variance appears in this regime as well.

The final three spectra  (letters G--I in Figures~\ref{fig:RMSErrors}--\ref{fig:SelectSpec}, and the bottom row of panels in Figure~\ref{fig:SelectSpec}) investigate the relatively poor performance of our prescription at $z \sim 2$ on small scales and lower number densities. Note first that the RMS error is in these cases still less than 5 per cent (Figure~\ref{fig:RMSErrors}), although ST performs much better (1--2 per cent). Note also that at $z \sim 2$, smoothing on this scale puts one at or past the Poisson limit for the specified number density. Nevertheless, it is interesting that inspection of the spectra (Figure~\ref{fig:SelectSpec}) shows the same higher-than-predicted power at intermediate scales, and the low overall per cent error comes from the predominance of the well-fit higher $k$-modes.\footnote{Note that the ``wiggles'' in these spectra (due to the finite volume of the Millennium Simulation) remain essentially identical from $z \sim 2$ to $z = 0$, reflecting the fact that linear growth uniformly augments the amplitudes of large-scale simulation modes without rearranging them.}

We therefore conclude that our prescription is typically accurate to within 5 per cent and is often comparable to that of the ST prescription for the galaxy power spectrum. Nevertheless, there is potentially room for improvement at scales around $10h^{-1}$ Mpc. In addition, the scale of the Millennium Simulation and the resulting cosmic variance makes it difficult to obtain a precise estimate of the accuracy of our prescription on larger scales -- although we note that on the largest, linear scales, the distribution is sufficiently Gaussian to obviate the need for sufficient statistics such as $A^*$. Larger-volume high-resolution simulations would however permit a better assessment of the large-scale accuracy of our prescription.

\section{Conclusion}
\label{sec:concl}
As noted in the introduction, the optimal observable for galaxy surveys is not the overdensity $\delta_g = N/\overline{N}-1$, but rather $A^*(N)$ -- because the power spectrum of the alternate statistic $A^*$ avoids the information plateau that besets the standard power spectrum $P_g(k)$ at small scales. However, in order to realize the potential of $A^*$, one must have in place a prescription for $P_{A^*}(k)$ with which to compare survey results.

We have shown that $A^*$ decomposes naturally into a continuous part $\tilde{A}$ and a stochastic part $\delta\!A^*$, and that the power spectrum decomposes in a similar manner. The contribution of the stochastic part is a discreteness plateau (Equation~\ref{eq:PkdeltAstar}) similar to the $1/\overline{n}$ shot noise plateau in the standard power spectrum. For the continuous part, we find that one can obtain the power spectrum of $\tilde{A}$ from the dark matter log spectrum $P_A(k)$ via an amplitude shift (the bias $b_{A^*}^2$) and a shape change parametrized by quantities $D$ and $B$ (as long as we restrict our consideration to scales at which the survey is not shot-noise dominated). We have also provided prescriptions for each of these quantities.

In addition, we have tested our prescription for $P_{A^*}(k)$ using discrete realizations of the Millennium Simulation and its rescalings; we find a typical accuracy around 5 per cent, although the value fluctuates depending on scale, redshift, etc. This accuracy is in most cases better than 5 per cent and is in general comparable to that of the (standard) ST prescription utilized in \textsc{Camb} for the nonlinear power spectrum.

We thus now have a procedure for predicting the power spectrum and mean of the discrete sufficient statistic $A^*$ for near-concordance cosmologies. As we and our collaborators show in previous work, this prediction is necessary in order to make full use of the data to be returned by future surveys. In particular, as \citet{WCS2015a,WCS2015b} have shown, the use of $A^*$ rather than the standard power spectrum can at a stroke double the information gleaned from such surveys.

Our prescription for predicting $P_{A^*}(k)$ is thus a major component of an approach that could result in a non-incremental multiplication of the effectiveness of WFIRST and \textit{Euclid}. Besides possible improvement of this prescription, remaining work includes analysis of the effects of both redshift space distortions and galaxy bias upon the power spectrum of $A^*$. The resultant information multiplication has the potential to advance our ultimate goal of characterizing the Universe.

\section*{Acknowledgements}
The Millennium Simulation data bases used in this Letter and the web application providing online access to them were constructed as part of the activities of the German Astrophysical Virtual Observatory (GAVO). This work was supported by NASA Headquarters under the NASA Earth and Space Science Fellowship program -- ``Grant 80NSSC18K1081'' -- and AR gratefully acknowledges the support. IS acknowledges support from National Science Foundation (NSF) award 1616974. 

\bibliography{PkAstar}

\appendix
\renewcommand{\theequation}{A\arabic{equation}}
\setcounter{equation}{0}

\section*{Appendix}
This appendix contains demonstrations of certain results which we quote in the main text.

First, we prove that the field $\delta\!A^*(\mathbf{r})$ is uncorrelated, where
\begin{equation}
\delta\!A^*(\mathbf{r}) = A^*(\mathbf{r}) - \tilde{A}\left(A(\mathbf{r})\right).
\label{eq:defdeltA}
\end{equation}

We first note that the mean of $\delta\!A^*$ vanishes, since
\begin{align}
\langle A^* \rangle & = \sum_{N=0}^\infty A^*(N) \int dA \,\mathcal{P}(A) \mathcal{P}(N|A) \\
	& = \int dA \left(\sum_{N=0}^\infty \mathcal{P}(N|A) A^*(N) \right) \mathcal{P}(A) \\
	& = \langle \tilde{A} \rangle,
\label{eq:eqmeans}
\end{align}
the last equality due to Equation~\ref{eq:Atilde}.

We now let $\xi_f$ denote the two-point correlation function of a field $f$, so that $\xi_f(r) = \langle f(\mathbf{x}) f(\mathbf{x+r}) \rangle - \langle f \rangle^2$; similarly, we let $\xi_{f\!f'}$ denote the cross-correlation of two fields, so that $\xi_{f\!f'}(r) = \langle f(\mathbf{x}) f'(\mathbf{x+r}) \rangle - \langle f \rangle \langle f' \rangle$. From Equations~\ref{eq:defdeltA} and \ref{eq:eqmeans} it follows that
\begin{equation}
\xi_{\delta\!A^*}(r) = \xi_{A^*}(r) - 2\xi_{A^*\!\tilde{A}}(r) + \xi_{\tilde{A}}(r).
\end{equation}
To show that $\delta\!A^*$ is uncorrelated, it thus suffices to demonstrate that $\xi_{A^*\!\tilde{A}}(r) = \xi_{A^*}(r) = \xi_{\tilde{A}}(r)$.

For the first equality,
\begin{align}
\xi_{A^*\!\tilde{A}}(r) & = \left\langle A^*(\mathbf{x}) \tilde{A}(\mathbf{x+r}) \right\rangle\\
& = \sum_{N_1} A^*(N_1) \int dA_2 \,\mathcal{P}(N_1, A_2) \tilde{A}(A_2).\label{eq:jprobint}
\end{align}
Here subscripts 1, 2 refer (respectively) to the values of the field at a given point $\mathbf{x}$ and at another point $\mathbf{x}+\mathbf{r}$; thus $\mathcal{P}(N_1, A_2)$ is the joint probability of finding a number count $N$ at point $\mathbf{x}$ and a log matter overdensity $A$ at point $\mathbf{x+r}$. Using Equation~\ref{eq:Atilde} to expand $\tilde{A}$, the integral in Equation~\ref{eq:jprobint} becomes
\begin{align}
\lefteqn{\int dA_2 \sum_{N_2} A^*(N_2) \mathcal{P}(N_1|A_2) \mathcal{P}(N_2|A_2) \mathcal{P}(A_2)}\nonumber\\
	& = \sum_{N_2} A^*(N_2)\int dA_2 \,\mathcal{P}(N_1|A_2) \mathcal{P}(A_2|N_2) \mathcal{P}(N_2)\label{eq:useBayes} \\
	& = \sum_{N_2} A^*(N_2)\mathcal{P}(N_1|N_2) \mathcal{P}(N_2),\label{eq:useMarkov}
\end{align}
where Equation~\ref{eq:useBayes} follows from Bayes' Theorem, and Equation~\ref{eq:useMarkov} follows from the fact that $N_1$ depends on $N_2$ only through $A_2$ (i.e., the number counts are correlated only because the underlying dark matter is correlated). Combining Equations~\ref{eq:jprobint} and \ref{eq:useMarkov} we thus have
\begin{align}
\xi_{A^*\!\tilde{A}}(r) & = \sum_{N_1,N_2} \mathcal{P}(N_1, N_2) A^*(N_1) A^*(N_2)\\
& = \left\langle A^*(\mathbf{x}) A^*(\mathbf{x+r}) \right\rangle = \xi_{A^*},
\end{align}
which was to be proved.

A similar argument shows that $\xi_{A^*}(r) = \xi_{\tilde{A}}(r)$:
\begin{align}
\xi_{\tilde{A}}(r) & = \left\langle \tilde{A}(\mathbf{x}) \tilde{A}(\mathbf{x+r}) \right\rangle\\
& = \int dA_1\,dA_2 \mathcal{P}(A_1,A_2) \tilde{A}(A_1) \tilde{A}(A_2)\\
\begin{split}
	& = \sum_{N_1, N_2} A^*(N_1) A^*(N_2)\; \times \\
	& \hspace{0.7cm}\int dA_1\,dA_2 \mathcal{P}(A_1|A_2) \mathcal{P}(A_2) \mathcal{P}(N_1|A_1)\mathcal{P}(N_2|A_2)
\end{split}\\
\begin{split}
	& = \sum_{N_1, N_2} A^*(N_1) A^*(N_2)\; \times \\
	& \hspace{0.7cm} \int dA_1\,dA_2 \mathcal{P}(A_1|A_2) \mathcal{P}(N_1|A_1)\mathcal{P}(A_2|N_2)\mathcal{P}(N_2)
\end{split}\\
& = \sum_{N_1, N_2} A^*(N_1) A^*(N_2) \mathcal{P}(N_1|N_2)\mathcal{P}(N_2)\\
& = \left\langle A^*(\mathbf{x}) A^*(\mathbf{x+r}) \right\rangle = \xi_{A^*}.
\end{align}
It follows that the $\delta\!A^*$ field is uncorrelated, so Equation~\ref{eq:PAsdisc} yields its power spectrum.

Second, we note that since $\xi_{A^*}(r) = \xi_{\tilde{A}}(r)$, and since $A^* = \tilde{A} + \delta\!A^*$, we can say that
\begin{equation}
P_{A^*}(k) = P_{\tilde{A}}(k) + P_{\delta\!A^*}.
\end{equation}

Finally, we can obtain an expression for $\sigma^2_{\delta\!A^*}$, by first considering $\left\langle A^* \tilde{A} \right\rangle$:
\begin{align}
\left\langle A^* \tilde{A} \right\rangle & = \int dA \sum_N \mathcal{P}(N, A) A^*(N) \,\tilde{A}(A)\\
	& = \int dA\, \mathcal{P}(A) \tilde{A}(A)\cdot \sum_N \mathcal{P}(N|A) A^*(N)\\
	& = \int dA\, \mathcal{P}(A)\, \tilde{A}(A)^2\\
	& = \left\langle \tilde{A}(A)^2 \right \rangle.
\end{align}

Since $\langle \delta\!A^* \rangle$ vanishes by Equation~\ref{eq:eqmeans},
\begin{align}
\sigma^2_{\delta\!A^*} & = \left\langle \left(\delta\!A^*\right)^2 \right\rangle\\
	& = \left\langle (A^*)^2 \right\rangle - 2\left\langle A^* \tilde{A} \right\rangle + \left\langle \tilde{A}^2 \right\rangle\\
	& = \left\langle (A^*)^2 \right\rangle - \left\langle \tilde{A}^2 \right\rangle\\
	& = \sigma^2_{A^*} - \sigma^2_{\tilde{A}}.
\end{align}

\label{lastpage}
\end{document}